\documentclass[submit]{epsv8}

\usepackage{amsmath,amssymb}
\usepackage{graphicx}
\usepackage{siunitx}
\usepackage{booktabs}
\usepackage[hidelinks]{hyperref}
\graphicspath{{./}}

\renewcommand{\theappendix}{\Alph{appendix}}   %
\makeatletter   %
\renewcommand{\paragraph}{\@startsection{paragraph}{4}{0pt}{0.5\baselineskip}{-1em}{\normalfont\normalsize\bfseries}}
\makeatother
\title{C$_2$ Swan, CH and CN emission, surface temperature and luminous efficiency of the Hayabusa2 sample-return capsule from ground-based spectroscopy}
\author{
Shinsuke Abe$^{1,2*}$ (abe.shinsuke@nihon-u.ac.jp),
Hadrien A. R. Devillepoix$^{3,4}$,
Eleanor K. Sansom$^{3,4}$,
Masa-yuki Yamamoto$^{5}$,
Satoshi Tanaka$^{6}$,
Yoshiaki Ishihara$^{7}$,
Kazuhisa Fujita$^{8}$,
Makoto Yoshikawa$^{6}$,
Martin C. Towner$^{3}$,
Martin Cup\'ak$^{3}$,
Kei-ichi Okuyama$^{1,2}$,
Takumi Ogawa$^{1}$,
Amane Uda$^{1}$,
Satoshi Nozawa$^{9}$,
Yoshihiro Hiramatsu$^{10}$,
Taichi Kawamura$^{11}$,
Norihisa Segawa$^{12}$,
Yoshihiro Kakinami$^{13}$,
Andrew Cool$^{14}$,
Geoffrey Bonning$^{15}$,
Hirohide Demura$^{16}$\\[1ex]
$^{1}$Department of Aerospace Engineering, Nihon University, 7-24-1 Narashinodai, Funabashi, Chiba, 274-8501, Japan\\
$^{2}$Space Science Research Unit (NU-SX), Research Institute of Science and Technology, Nihon University, 1-8-14 Kanda Surugadai, Chiyoda-ku, Tokyo, 101-8308, Japan\\
$^{3}$Space Science and Technology Centre, Curtin University, Kent Street, Bentley, Perth, Western Australia, 6102, Australia\\
$^{4}$Curtin Institute for Radio Astronomy, Curtin University, Perth, Western Australia, Australia\\
$^{5}$School of Systems Engineering, Kochi University of Technology, 185 Miyanokuchi, Tosayamada, Kami, Kochi, 782-8502, Japan\\
$^{6}$Institute of Space and Astronautical Science, Japan Aerospace Exploration Agency, 3-1-1 Yoshinodai, Chuo-ku, Sagamihara, Kanagawa, 252-5210, Japan\\
$^{7}$Human Spaceflight Technology Directorate, Japan Aerospace Exploration Agency, 2-1-1 Sengen, Tsukuba, Ibaraki, 305-8505, Japan\\
$^{8}$ElevationSpace Inc., Ichigo Kakyoin Building 9F, 2-1-65 Kakyoin, Aoba-ku, Sendai, Miyagi, 980-0013, Japan\\
$^{9}$Faculty of Science, Ibaraki University, 2-1-1 Bunkyo, Mito, Ibaraki, 310-8512, Japan\\
$^{10}$Faculty of Geosciences and Civil Engineering, Institute of Science and Engineering, Kanazawa University, Kakuma-machi, Kanazawa, Ishikawa, 920-1192, Japan\\
$^{11}$Institut de Physique du Globe de Paris, 35 Rue H\'el\`ene Brion, Paris, 75013, France\\
$^{12}$Faculty of Information Science and Engineering, Kyoto Sangyo University, Kamigamo-Motoyama, Kita-ku, Kyoto, 603-8555, Japan\\
$^{13}$Space Information Center, Hokkaido Information University, 59-2 Nishi-nopporo, Ebetsu, Hokkaido, 069-0832, Japan\\
$^{14}$The River Murray International Dark Sky Reserve, PO Box 10, Cambrai, South Australia, 5353, Australia\\
$^{15}$Research School of Earth Sciences, The Australian National University, Building 142, 1 Mills Road, Acton, Canberra, 2601, Australia\\
$^{16}$Aizu Research Center for Space Informatics (ARC-Space), The University of Aizu, Tsuruga, Ikki-machi, Aizuwakamatsu, Fukushima, 965-8580, Japan\\[1ex]
$^*$Corresponding author: abe.shinsuke@nihon-u.ac.jp
}
\abstract{
The Hayabusa2 sample-return capsule entered the atmosphere over South Australia on 5 December 2020 at
\SI{12}{km\,s^{-1}}. A slitless spectral video, taken from the ground with a hand-tracked consumer camera,
shows the capsule alone from 80 to 49~km. In the first ten seconds of
heating the shock layer emits the CN violet system, the CH A--X band and the C$_2$ Swan bands, the latter for the first
time in remote observations of a capsule entry. C$_2$ ($T_{\rm rot}\approx\SI{9200}{K}$,
$T_{\rm vib}\approx\SI{7800}{K}$) and CN (about \SI{7000}{K}) are far hotter than the surface. C$_2$ and CH
disappear while the heat flux is still rising and CN persists, which we relate to the early pyrolysis of the
carbon-phenolic ablator and to the changing boundary-layer chemistry. The colour temperature of the heat shield
rises from 2720 to \SI{3080}{K}, in agreement with airborne results. The radiated energy is one to two orders of
magnitude below that of natural fireballs, because the light comes from the hot surface rather than from ablation
vapour. A faint persistent train lights up about \SI{1}{s} after the passage and fades within
a few seconds, probably chemiluminescence in the cooling wake.
}
\begin{document}
\renewcommand{\today}{November 12, 2026}
\maketitle

\section{Introduction}
\label{sec:intro}

The sample-return capsule (SRC) of the Hayabusa2 mission \citep{Watanabe2017}, carrying material from
asteroid Ryugu \citep{Yada2022}, entered the atmosphere over South Australia at about 17:28~UTC on
5 December 2020 at about \SI{12}{km\,s^{-1}} and was recovered in the Woomera range
\citep{Yamada2021,YamadaYoshihara2022}. Like its predecessor Hayabusa in 2010, the capsule is a small
(\SI{0.41}{m} diameter) blunt body protected by a carbon-fibre-reinforced plastic (CFRP) ablator
\citep{YamadaYoshihara2022}. The night-time entry was observed by a coordinated international campaign
with ground-based cameras, spectrographs, infrasound and seismic sensors \citep{Sansom2022,Nishikawa2022},
and by two instrumented aircraft \citep{Scott2022,Zander2021}.

Optical spectra of a capsule during entry give two complementary diagnostics. The continuum, thermal emission
of the hot heat shield, gives its temperature history, which can be compared with post-flight
aerothermal and ablation analyses \citep{Suzuki2014,Winter2019}. The emission lines and molecular bands
show the shock layer and the ablation products; for carbon-based ablators the CN violet system is the
most prominent band, and its rotational and vibrational temperatures probe the gas near the
surface \citep{Jenniskens2010CN,Kraetzig2014}. Such measurements were made from aircraft and from the
ground for the Stardust capsule in 2006 \citep{Jenniskens2010echelle,Jenniskens2010CN,WinterTrumble2011,Harms2010,Stenbaek2010}
and for Hayabusa in 2010
\citep{Fujita2011PASJ,Ohnishi2011,Abe2011,Borovicka2011,Sato2012,Buttsworth2013,Kraetzig2014,Lohle2013}.
In the Hayabusa case, the bright fragmenting spacecraft bus dominated the spectra above about 45~km, and
the capsule continuum could be isolated only at lower altitudes \citep{Abe2011,Ohnishi2011}.

For Hayabusa2, which had no accompanying bus fragments, surface temperatures have so far been derived from
the airborne observations: single-temperature Planck fits gave peak values of about \SI{2760}{K}
\citep{Birch2022a} and \SI{3200}{K} after correction for atmospheric transmission \citep{Birch2022b}, a
surface-temperature distribution model gave \SIrange{2500}{3700}{K} \citep{Ravichandran2025}, and grism
spectra were analysed by \citet{Liu2025}; the airborne spectra are dominated by atomic N, O and
H lines of the shock layer \citep{Tanno2022,Birch2022b,Ravichandran2025}. The ground campaign \citep{Sansom2022} included a slitless spectral video camera at Coober Pedy.

In this paper we present the first detailed quantitative analysis of that slitless video spectrum. Because the capsule entered without a disintegrating spacecraft bus,
the spectrum shows the capsule alone from 80~km downwards, i.e.\ from the very beginning of the heating pulse.
We address four questions. (i) Which hot species does the shock layer of a carbon-phenolic ablator emit in the
earliest phase of the entry, and at which temperatures? We find, besides CN, the C$_2$ Swan bands and the CH
band, which have not been reported in remote observations of a sample-return capsule, and follow their disappearance near 69~km. (ii) How
do the colour and brightness temperatures of the heat shield evolve, and how do they compare with the
airborne results and the predictions? (iii) What fraction of the kinetic energy is radiated, i.e.\ what is the
luminous efficiency of a body of known mass and shape at about \SI{10}{km\,s^{-1}}, a quantity central to meteor
physics \citep{Abe2009LNP}? (iv) Does the capsule have a luminous wake, or leave a persistent train, as meteors do? After the observation and the data reduction
(Sections~\ref{sec:obs} and \ref{sec:reduction}; details in the appendices), we present the emission of the gas (molecular bands
and atomic lines), the surface temperature, the luminous efficiency and the heating, and the wake and train
(Section~\ref{sec:results}), and discuss them with the ground tests of carbon ablators, the Hayabusa and Stardust
entries and meteors (Section~\ref{sec:discussion}); the limitations of the method are summarised in
Section~\ref{sec:limits}.

\section{Observation}
\label{sec:obs}

The entry was recorded by a ground team of Curtin University (H. A. R. Devillepoix and E. K. Sansom) near Coober Pedy, South Australia
(Coober Pedy West, \ang{29.084}\,S, \ang{134.553}\,E), as part of the campaign described by \citet{Sansom2022}. The camera was a Sony
ILCE-7SM2 ($\alpha$7S~II), de-filtered (Bayer array with micro-lenses removed), with a Sigma 24~mm F1.4 DG~HSM Art lens and an Edmund Optics \#49584
600~lines/mm blazed transmission grating (50~mm square) in a rotatable holder in front of the lens.
The video was recorded in 3840$\times$2160 pixels, 8-bit data, at 25 frames per second with ISO~1600, F1.4 and a
1/4\,s exposure, so that every exposure is repeated in seven consecutive frames (3.6 exposures
per second). The per-frame Sony metadata (exposure, ISO, F-number and camera clock) were read from the
video stream. The 87 exposures of the event cover \SIrange{82.6}{48.1}{km} altitude; the capsule spectrum is measured in 72 of them,
between \SIrange{79.7}{49.2}{km} ($t=0.2$--21.4~s, 17:28:38.6 to 17:28:59.8~UTC). The image is monochrome (R=G=B) and the transfer curve was
identified as S-Log3 (Section~\ref{sec:tone}).
The waning gibbous Moon (age 20.5~d, 74\,\% illuminated) was above the horizon at an altitude of
\SI{33}{\degree} in the north-north-east (azimuth \SI{32}{\degree}), 53--\SI{87}{\degree} from the capsule
during the observation (\SI{53}{\degree} at peak brightness). It was outside the field of view; its light adds a
smooth sky background, which is removed with the dark level (Section~\ref{sec:dark}).

\begin{figure}[tbp]
\centering
\includegraphics[width=\textwidth]{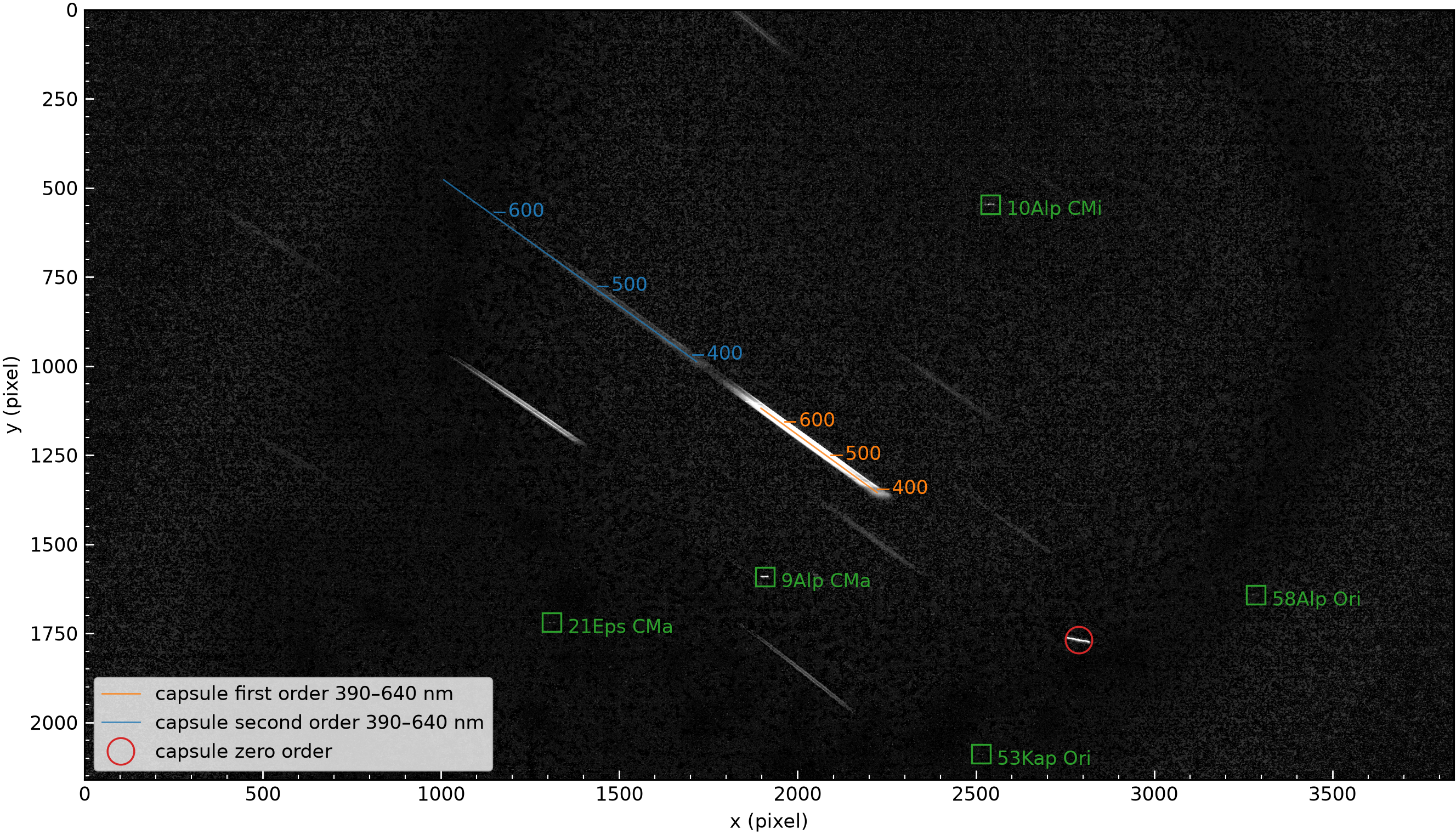}
\caption{\textbf{Slitless exposure of the Hayabusa2 capsule with the model spectral traces.} One 1/4\,s exposure ($t=10.8$~s, 17:28:49.2~UTC, 62.8~km altitude, 75~km range), linearised and
background-subtracted. The capsule zero order (red circle), its first- and second-order spectra
(model traces, \SIrange{390}{640}{nm}) and the identified stars are marked. The diffuse spectrum
crossing the field is Sirius, trailed because the camera followed the capsule.}
\label{fig:frame}
\end{figure}

The camera was moved by hand to follow the capsule. Its zero-order image is therefore a
curved streak of a few tens of pixels in each exposure, while the stars are trailed in the
exposures where the capsule is followed well (Figure~\ref{fig:frame}).

The image data of every exposure and the individual grey-body fits are shown in Appendix~\ref{app:exposures}.

\section{Data reduction}
\label{sec:reduction}

\subsection{Exposures, astrometry and timing}
Exposure boundaries were found from frame differences and confirmed by the metadata; each exposure is
the average of its frames after the first one, which is affected by the video compression. Plate
solutions in J2000 were obtained from bright stars (Sirius, Procyon, Betelgeuse and others) with a
gnomonic projection plus radial distortion (focal length \SI{2658.7}{px}, $k_1=-0.030$). Reliable
solutions exist for the exposures in which at least two stars are not trailed; they give the direction of
the capsule in J2000.
Comparison with the predicted entry trajectory (Section~\ref{sec:limits}) fixes the
observing site and the time offset of the video: the camera clock was \SI{321.4}{s} fast and video time
zero corresponds to 17:26:58.58~UTC. A fit with free site coordinates is consistent with the site within
0.7\,\si{\degree} (rms \SI{7.9}{arcmin}); with the adopted site the rms of the capsule directions is
\SI{11.4}{arcmin}. Altitude, range and elevation of every exposure follow from the intersection of the
observed direction with the trajectory. These quantities depend only on the trajectory geometry, in which the predicted and
best-estimated trajectories agree (Section~\ref{sec:limits}); the time epoch of the predicted trajectory affects only the UTC of each exposure.
Throughout this paper, time is given as $t$, the time after the capsule passed 80~km altitude on the trajectory
(17:28:38.4~UTC), which is independent of the absolute time epoch of the trajectory (Section~\ref{sec:limits}).

\subsection{Grating model and wavelength calibration}
A ray $\mathbf{u}$ in camera coordinates is diffracted by the grating in front of the lens as
\begin{equation}
\mathbf{v}_{xy} = \mathbf{u}_{xy} + \frac{m\lambda}{d}\,\mathbf{g},
\end{equation}
where $m$ is the order, $d$ the grating period and $\mathbf{g}$ the unit grating vector (conical diffraction),
and the diffracted ray is projected with the lens model. This describes the dispersion and the curvature
of the spectra over the whole \SI{74}{\degree} field. The orientation of $\mathbf{g}$
(\SI{215.93}{\degree}, tilt $<\SI{0.01}{\degree}$) was fitted to 31 star traces (rms \SI{2.4}{px} across
the trace), and the period ($d=\SI{1670.1}{nm}$, 1.0020 times the nominal value) and a zero-point offset of
\SI{4.2}{px} to the Balmer lines of Sirius (residuals $\leq\SI{0.11}{nm}$). Na~D and H$\beta$ of the capsule
agree within \SI{0.5}{nm} in the first and second orders.

\subsection{Dark level, background and flat field}
\label{sec:dark}
No dedicated dark frames (lens cap on) were recorded, and the dark level is removed in four steps. (i) The
video black level (pixel value $=16$) is subtracted when the transfer curve is inverted
(Section~\ref{sec:tone}), i.e. in linear units. (ii) In each linearised exposure, a smooth background
surface, containing the sky and the spatially slowly varying dark signal, is subtracted: sources (spectra,
zero orders, stars) are found on a high-pass image and masked, and the background is the median of the
remaining pixels in $16\times16$-pixel blocks, smoothed. The image is then divided by a flat field (vignetting)
derived from the sky in the exposures after the fireball. (iii) During the extraction
(Section~\ref{sec:extraction}) a local background is taken with the same kernel shifted perpendicular to the
dispersion on both sides and subtracted. (iv) A residual zero level is measured at first-order positions
between the zero order and \SI{370}{nm} (\SIrange{300}{360}{nm}), which receive no light from any order, and
subtracted as a constant; it is below 1.1\,\% of the peak signal for 90\,\% of the exposures (median
$-0.02\,\%$; the largest, $-6.0\,\%$, in the first, faintest exposure at 79.7~km). The fixed-pattern dark of
individual pixels (hot pixels) is not removed; its effect is reduced by the weighted sum over many pixels in the
extraction.

\subsection{Matched extraction}
\label{sec:extraction}
During an exposure the capsule moves by tens of pixels at an angle to the dispersion, which would smear a
simple extraction by about \SI{30}{nm}. We therefore use the zero-order image $b(\mathbf{p})$ (normalised to
unit sum) as a two-dimensional kernel and sum the first-order image $I$ at the dispersed position of every
zero-order pixel,
\begin{equation}
M(\lambda) = \sum_j b_j\, I\big(\mathbf{p}_j + \mathbf{D}_m(\lambda)\big),
\end{equation}
where $\mathbf{D}_m$ is the displacement given by the grating model. $M$ equals the source spectrum
convolved with the autocorrelation of $b$ along the dispersion direction, which is computed for each
exposure and serves as its line-spread function (median FWHM \SI{2.7}{nm} in the first order and
\SI{0.83}{nm} in the second). This autocorrelation does not include the difference between the point-spread
function at the zero-order and at the dispersed position; for the second order an additional instrumental
width is determined from H$\gamma$ (Appendix~\ref{sec:order2}). The local background and the zero level are
subtracted as described in Section~\ref{sec:dark}. The absolute level is tied to a wide-aperture extraction across the
trace by one scale factor per exposure.

\subsection{Transfer curve}
\label{sec:tone}
Without linearisation, the ratio of first-order to zero-order signal of the capsule scatters by 34\,\%
between exposures and is correlated with how compact the zero-order image is, a signature of a compressive
transfer curve. Among the candidate curves (none, Rec.~709~\citealp{ITU709}, gamma 2.2, S-Log2, S-Log3) the S-Log3
inverse reduces this scatter to 10\,\%, and S-Log3 is adopted; ISO~1600 is also the lowest ISO of the
S-Log modes of this camera. The other curves are carried through the analysis as systematic variants.

\subsection{Spectral sensitivity and radiant intensity}
Sirius appears in seven of the exposures. Its aperture spectra divided by the CALSPEC spectrum
\texttt{sirius\_stis\_005} \citep{Bohlin2014}, convolved with each exposure's smear kernel and attenuated
for the extinction at its elevation (\SI{74.5}{\degree}), give the sensitivity curve. The relative
sensitivity peaks near \SI{510}{nm} and falls to \SI{40}{\%} at \SI{608}{nm} and to 20\,\% at
\SI{625}{nm} (infrared-cut filter); we use the spectra up to \SI{640}{nm}. The flux of Sirius scatters by
18\,\% between exposures (scintillation through the \SI{17}{mm} aperture), and the absolute scale is the
mean of the seven exposures (uncertainty 7\,\%). Capsule spectra are corrected for extinction at the
capsule elevation (aerosol optical depth 0.08 assumed) and converted to radiant intensity with the range.

The second order has twice the dispersion but cannot be calibrated with Sirius. Its relative sensitivity
was obtained from the capsule itself, as the ratio of second-order counts to the calibrated first-order
intensity in 21 continuum-dominated exposures (\SIrange{60}{52}{km}) with emission features masked.

\subsection{Further corrections}
Four further steps are described in the appendices: the wavelength scale of each exposure, corrected for the
offset between the zero-order and dispersed images caused by the camera motion (Appendix~\ref{sec:wavecorr});
the resolution of the second order (Appendix~\ref{sec:order2}); the stray light, which is negligible
(Appendix~\ref{sec:stray}); and a signal-proportional correction of the sensitivity at the edges of the pass band,
where the infrared-cut filter edge depends on the field angle (Appendix~\ref{sec:selfcal}).

\section{Results}
\label{sec:results}

\subsection{Overview of the spectra}
\begin{figure}[tbp]
\centering
\includegraphics[width=\textwidth]{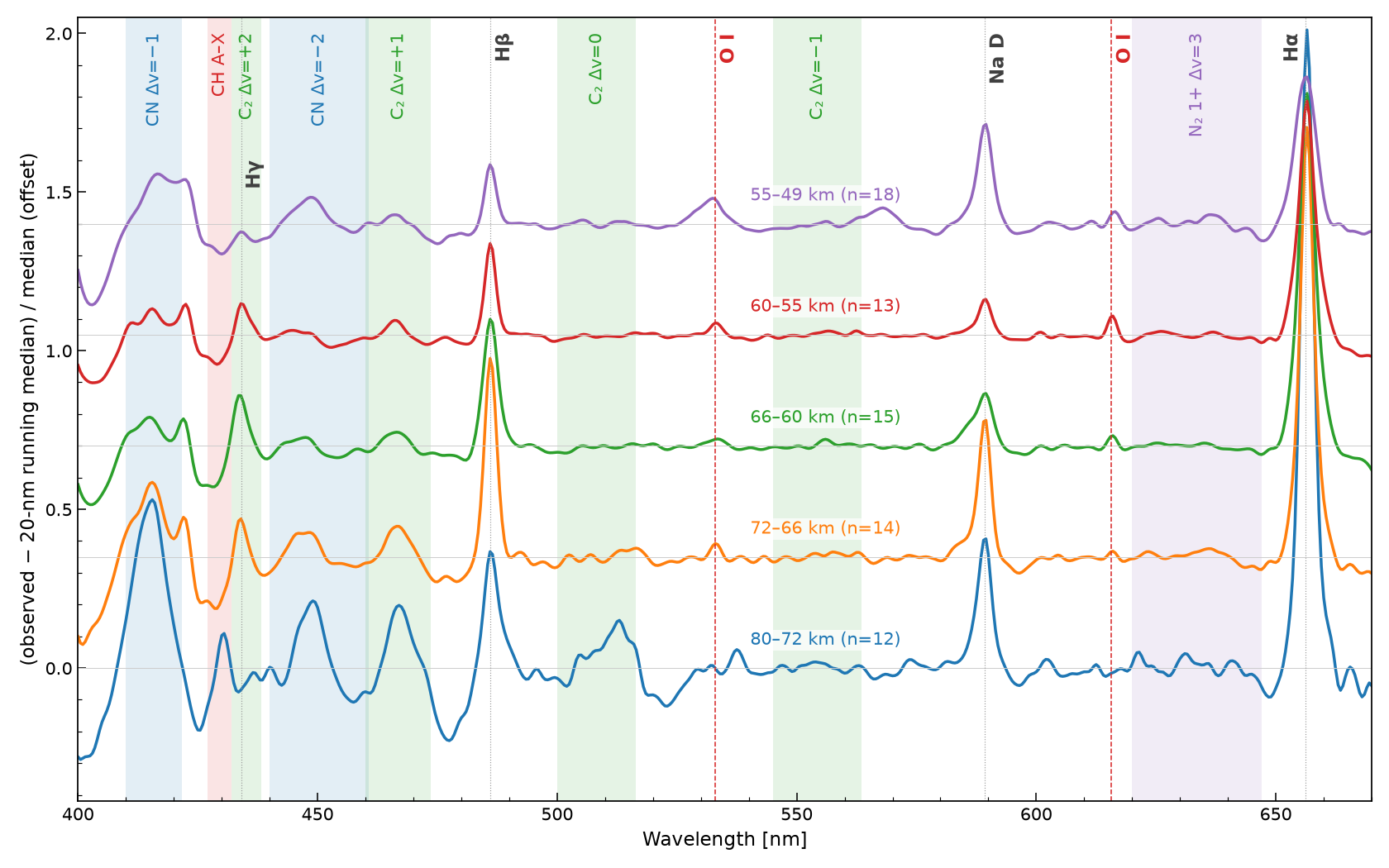}
\caption{\textbf{Emission features of the capsule spectrum in five altitude bins.} Emission features in five altitude bins: mean calibrated first-order spectrum minus a \SI{20}{nm}
running median, divided by the running median (offset for clarity; highest altitude at the bottom). The
wavelength scale is corrected for each exposure and for the mean residual offset (Appendix~\ref{sec:wavecorr}).
Shaded: molecular band sequences (blue: CN B--X $\Delta v=-1$ and $-2$; red: CH A--X (0,0); green: C$_2$ Swan
$\Delta v=+2$ to $-1$; violet: N$_2$ first positive $\Delta v=3$, inside the range of the
signal-proportional correction of Appendix~\ref{sec:selfcal}). Grey dotted: identified atomic lines (H$\gamma$,
H$\beta$, Na~D, H$\alpha$); red dashed: O~I 533.0 and \SI{615.7}{nm} (Section~\ref{sec:lines}).}
\label{fig:altspec}
\end{figure}

The radiant intensity in \SIrange{400}{680}{nm} peaks at \SI{4.8e3}{W\,sr^{-1}} at $t=12.2$~s (17:28:50.6~UTC) and
60.8~km altitude (range 66~km), corresponding to an apparent V magnitude of $-5.0$ and an absolute
magnitude (100~km) of $-4.1$; the S-Log2 and Rec.~709 transfer curves give $-4.7$ and $-4.5$ (absolute $-3.8$ and
$-3.6$).
Figure~\ref{fig:altspec} shows the emission features of the mean spectra in altitude bins, after removal of
the continuum by a running median. The CN violet system ($\Delta v=-1$ at \SIrange{415}{421.6}{nm}; $\Delta v=0$ at
\SI{388.3}{nm} lies below the plotted range), H$\gamma$, H$\beta$,
H$\alpha$ (at the edge of the sensitivity), Na~D and O~I 533.0 and \SI{615.7}{nm} are seen, and at 80--72~km the CH
A--X band at \SI{431}{nm} (Section~\ref{sec:ch}). The CN bands are relatively strongest at high altitude. The
structure at 612--\SI{616}{nm}, present also in the uncalibrated spectra, is partly instrumental (edge of the
infrared-cut filter, Appendix~\ref{sec:selfcal}) and partly O~I emission.

\subsection{Shock-layer molecular emission: CN, CH and C$_2$ Swan}
\label{sec:molecules}
The bright, high-temperature molecular emission of the shock layer is the most distinctive feature of the
capsule spectrum at high altitude (Figure~\ref{fig:altspec}): the CN violet system dominates the near
ultraviolet, and above 69~km the C$_2$ Swan bands and the CH A--X band appear. All three are modelled with
METEORAD line-by-line spectra (Section~\ref{sec:cn}, Appendix~\ref{app:temps}).

\subsubsection{CN violet system}
\label{sec:cn}
The CN B$^2\Sigma^+$--X$^2\Sigma^+$ $\Delta v=-1$ sequence (\SIrange{400}{440}{nm}) was fitted in both
orders simultaneously with line-by-line spectra from METEORAD, a Python port of SPRADIAN \citep{FujitaAbe1997,FujitaAbe2003}
that we extended for meteor and debris spectra with neutral and ionized metal atoms (Na, K, Li, Ca, Fe, Cr, Mn, Mg, Al,
Si, Ti, Ni, Cu) and the oxides AlO, TiO, FeO and MgO (for CN, N$_2^+$ and C$_2$ its spectra are identical to
SPRADIAN-1.8), computed for
optically thin emission on a grid of rotational and vibrational temperatures (\SIrange{1000}{16000}{K};
$T_{\rm tr}=T_{\rm rot}$, $T_{\rm ex}=T_{\rm vib}$), converted to air wavelengths and convolved with the
line-spread function of each order (first order \SI{2.7}{nm}; second order on the \SI{0.1}{nm} grid with
the resolution fitted per bin, Appendix~\ref{sec:order2}). The model also contains N$_2^+$ B--X (0,1) and (1,2), H$\delta$,
H$\gamma$, Ca~I \SI{422.7}{nm}, a linear continuum per order and a wavelength shift per order. The
$\Delta v=0$ sequence below \SI{390}{nm} lies where the sensitivity falls steeply and is not used.

\begin{figure}[tbp]
\centering
\includegraphics[width=\textwidth]{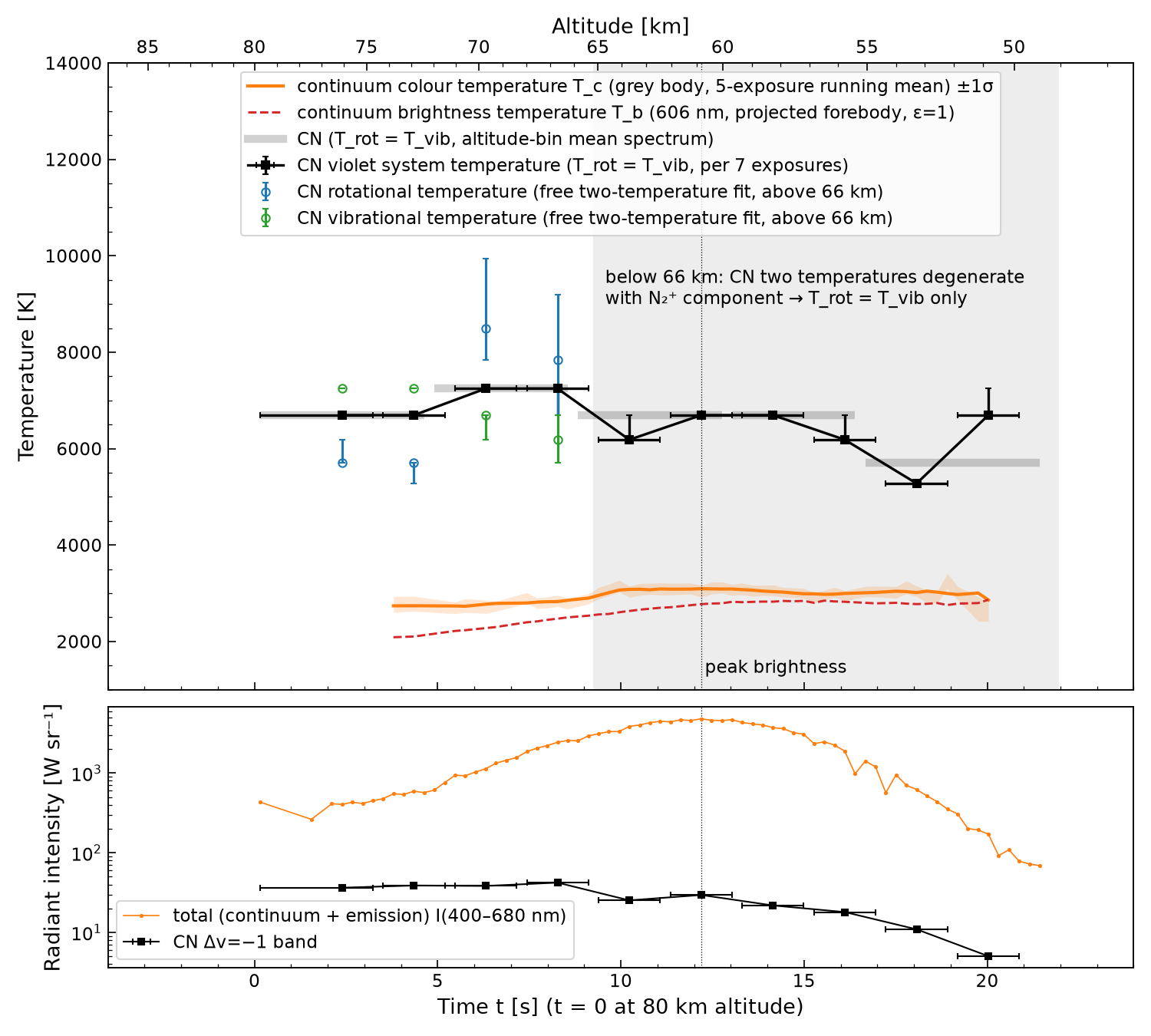}
\caption{\textbf{Temperature history of the heat-shield continuum and of CN.} Temperature history. Orange: continuum colour temperature (grey body, five-exposure
running mean, $\pm1\sigma$); red dashed: brightness temperature at \SI{606}{nm}; black: CN violet
temperature ($T_{\rm rot}=T_{\rm vib}$) in groups of seven exposures and (grey bars) in altitude bins; open
symbols: free $T_{\rm rot}$ and $T_{\rm vib}$ above 66~km. Bottom: total and CN $\Delta v=-1$ radiant
intensity. Time $t$ after the capsule passed 80~km, with the altitude along the trajectory on the top axis (as in
Figure~\ref{fig:heating}).}
\label{fig:history}
\end{figure}

\begin{figure}[tbp]
\centering
\includegraphics[width=\textwidth]{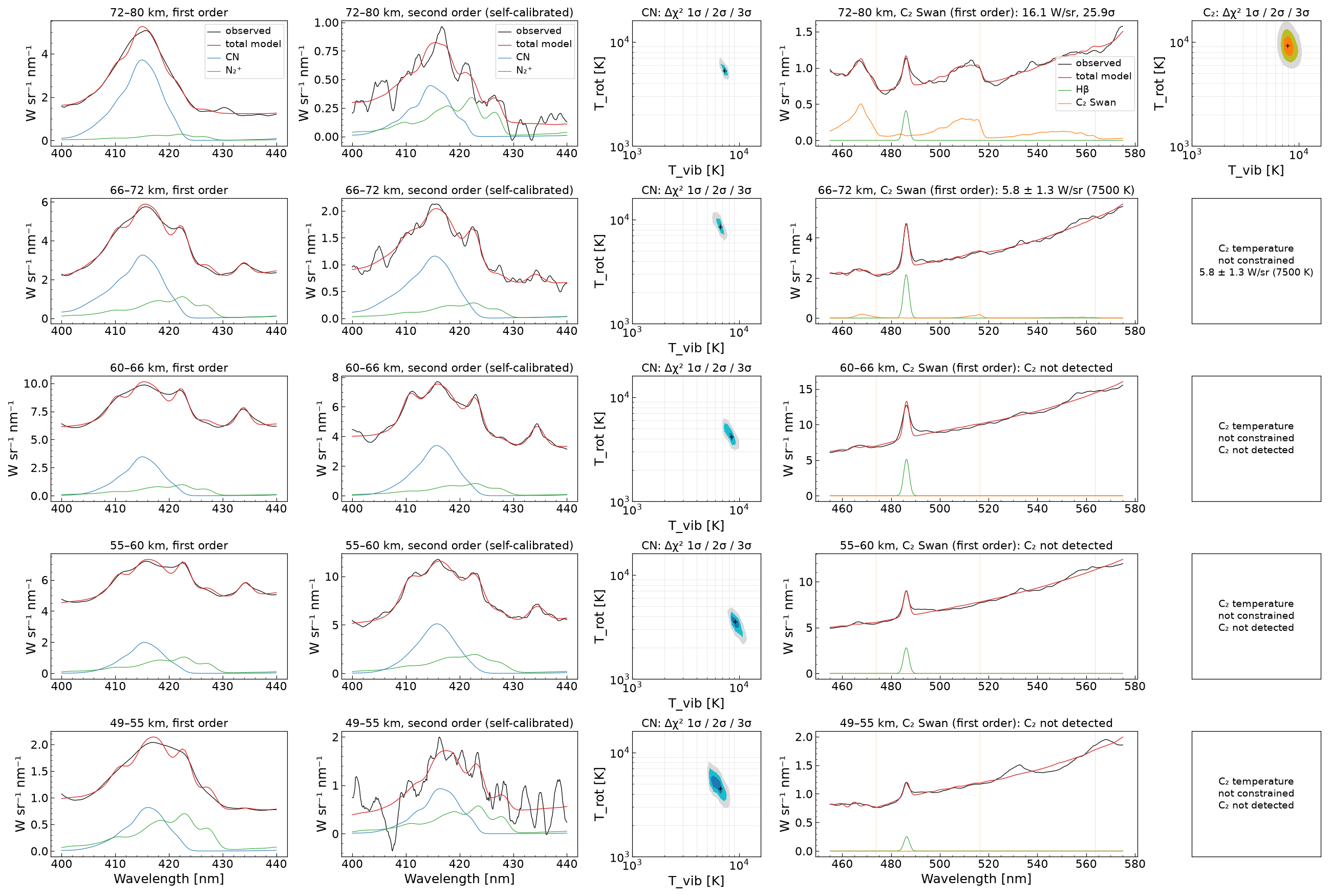}
\caption{\textbf{Line-by-line fits of the CN violet and C$_2$ Swan bands.} METEORAD fits in altitude bins. Columns 1--3: CN violet $\Delta v=-1$ and N$_2^+$ first negative
(0,1)/(1,2) at \SIrange{400}{440}{nm}; first order (\SI{2.7}{nm}) and second order on the \SI{0.1}{nm} grid
(self-calibrated; resolution fitted per bin: 2.0, 3.0, 3.0, 3.5, \SI{2.25}{nm}); black: observed; red: total
model; blue: CN; green: N$_2^+$ (components without the continuum); $\Delta\chi^2$ regions (1, 2, 3$\sigma$) of
$T_{\rm rot}$ and $T_{\rm vib}$ of CN. Columns 4--5: C$_2$ Swan at \SIrange{455}{575}{nm} (first order); red: total model (C$_2$, CN and N$_2^+$
$\Delta v=-2$, H$\beta$, continuum); green: H$\beta$; orange: C$_2$ Swan alone;
dotted: Swan $\Delta v=+1$, 0, $-1$ heads; $\Delta\chi^2$ regions of the C$_2$ temperatures where Swan is detected,
otherwise the band at \SI{7500}{K} where it is detected.}
\label{fig:molfits}
\end{figure}

Between 80 and 66~km, where CN dominates the \SIrange{400}{440}{nm} emission, the fit gives
$T_{\rm rot}=\SI{5300}{K}$ and $T_{\rm vib}=\SI{7200}{K}$ at 80--72~km and $T_{\rm rot}=\SI{8500}{K}$ and
$T_{\rm vib}=\SI{6700}{K}$ at 72--66~km
(Figure~\ref{fig:molfits}); changing the line-spread function by $\pm10$\,\%, the continuum to quadratic or the
window to \SIrange{403}{437}{nm} gives $T_{\rm rot}=\SIrange{5700}{9200}{K}$ (except \SI{4500}{K} with the
narrower window at 80--72~km and with the quadratic continuum at 72--66~km) and $T_{\rm vib}=\SIrange{6700}{8500}{K}$. Below 66~km a broad component attributed to N$_2^+$ at
$\sim\SI{10000}{K}$ competes with CN and the two temperatures become degenerate. Assuming
$T_{\rm rot}=T_{\rm vib}$, the CN temperature is \SIrange{6700}{7200}{K} in all altitude bins above 55~km and
\SI{5700}{K} at 55--49~km (\SIrange{5300}{7200}{K} in groups of seven exposures; Figure~\ref{fig:history}). The CN $\Delta v=-1$
radiant intensity of these fits (400--\SI{440}{nm}, groups of seven exposures) is about \SI{40}{W\,sr^{-1}} down to 66~km and decreases to \SI{5}{W\,sr^{-1}} at 51~km,
starting to decline before the peak of the total brightness.

\begin{figure}[tbp]
\centering
\includegraphics[width=\textwidth]{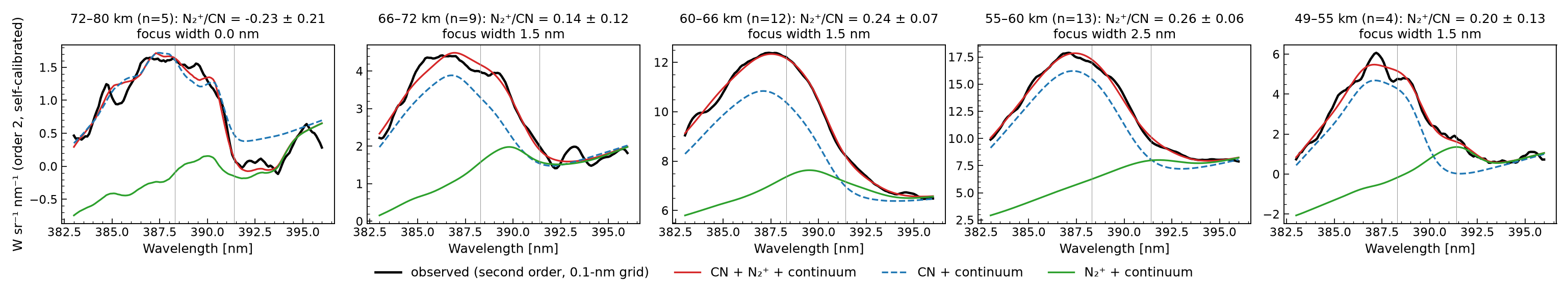}
\caption{\textbf{Second-order spectra of the CN and N$_2^+$ bands near 390~nm.} Second-order spectra (\SI{0.1}{nm} grid) at \SIrange{383}{396}{nm} in altitude bins. Black: mean of
the exposures fitted; red: mean of the per-exposure fits of CN $\Delta v=0$ plus N$_2^+$ (0,0)+(1,1) and a
linear continuum, each convolved with the exposure's measured line-spread function and the focus width of the
bin (given in the titles); blue dashed: the CN and continuum part of the same fits; green: the N$_2^+$ and continuum part. Vertical lines: CN (0,0) head at
\SI{388.3}{nm} and N$_2^+$ (0,0) head at \SI{391.4}{nm}. Titles: number of exposures, N$_2^+$(0,0)/CN($\Delta v=0$)
and the focus width.}
\label{fig:n2p}
\end{figure}

N$_2^+$ is formally significant in these fits, but it acts largely as a smooth component and its (0,1) head
at \SI{427.8}{nm} is not clearly seen; its (0,0) band is identified in the second order
(Section~\ref{sec:n2p}, Figure~\ref{fig:n2p}).

\subsubsection{CH A--X band}
\label{sec:ch}
At 80--72~km a band at 430--\SI{431}{nm}, beyond the
N$_2^+$ (0,1) head, is the CH A--X (0,0) band: adding a METEORAD CH spectrum (\SI{5000}{K}) to the first-order fit of
CN, N$_2^+$, H$\delta$, H$\gamma$ and Ca~I lowers $\chi^2$ by 61 and gives \SI{1.9 \pm 0.9}{W\,sr^{-1}}; the amplitude is zero below 72~km, so CH, like C$_2$, is
seen only at the highest altitudes. Fe~I \SI{431.7}{nm} is excluded because its stronger lines at 438.4 and
\SI{527.0}{nm} are absent.

\subsubsection{C$_2$ Swan system}
\label{sec:swan}
The C$_2$ d$^3\Pi_g$--a$^3\Pi_u$ Swan system was analysed in the same way in the first order at
\SIrange{455}{575}{nm}, which contains the $\Delta v=+1$ (\SI{473.7}{nm}), $\Delta v=0$ (\SI{516.5}{nm}) and
$\Delta v=-1$ (\SI{563.6}{nm}) sequences: METEORAD C$_2$ spectra on the $(T_{\rm rot},T_{\rm vib})$ grid, CN B--X $\Delta v=-2$ and N$_2^+$ B--X
$\Delta v=-2$ at the temperatures of the \SIrange{400}{440}{nm} fit (shapes fixed, amplitudes free), H$\beta$ and a
cubic continuum (Figure~\ref{fig:molfits}, right). At 80--72~km the structures at
467 and \SI{512}{nm} are reproduced by the Swan bands: with high rotational and vibrational excitation the
violet-degraded sequences (1,0)--(4,3) and (0,0)--(3,3) peak 4--\SI{6}{nm} short of the (0,0) heads at our
resolution. Swan is detected at 26$\sigma$ (\SI{16.1}{W\,sr^{-1}}, i.e. 0.43 of the CN $\Delta v=-1$ band) with
$T_{\rm rot}=\SI{9200}{K}$ (68\,\%: 7200--\SI{10800}{K}) and $T_{\rm vib}=\SI{7800}{K}$ (7200--\SI{8500}{K}); the
systematic variants (line-spread function $\pm10$\,\%, quadratic or quartic continuum, window
\SIrange{460}{560}{nm}) give $T_{\rm rot}=\SIrange{8500}{13700}{K}$, $T_{\rm vib}=\SIrange{5700}{8500}{K}$ and
\SIrange{14.1}{20.9}{W\,sr^{-1}}, and $T_{\rm rot}=T_{\rm vib}$ gives \SI{8500}{K}. Without the CN and N$_2^+$
components the band is stronger (\SI{19.1}{W\,sr^{-1}}) and the fit is worse (reduced $\chi^2$ 6.74 versus
5.35, mainly at 455--\SI{465}{nm}, where the CN $\Delta v=-2$ sequence lies), with $T_{\rm rot}=\SI{12600}{K}$
and $T_{\rm vib}=\SI{5700}{K}$.

The Swan emission ends near 69~km. With the temperatures of the 80--72~km fit, the band is
mostly 13--\SI{28}{W\,sr^{-1}} in single exposures at 77--71~km, 6--\SI{10}{W\,sr^{-1}} at 70.6--69.7~km
(3--7$\sigma$ each), $5.8\pm1.8$\,W\,sr$^{-1}$ at 69.3~km and consistent with zero below:
$9.9\pm1.0$\,W\,sr$^{-1}$ (10$\sigma$) for the mean of the seven exposures at
72--69~km and $3.3\pm2.3$\,W\,sr$^{-1}$ at 69--66~km. At 72--69~km the two temperatures cannot be fitted
independently (the free fit drifts to $T_{\rm rot}\approx\SI{2400}{K}$ because the band envelope at this
strength is degenerate with the continuum polynomial), but a single temperature $T_{\rm rot}=T_{\rm vib}$ gives
\SI{7200}{K} (68\,\%: 6200--\SI{9900}{K}), consistent with the 80--72~km fit with the same assumption
(\SI{8500}{K}) and with CN; at 69--66~km the temperature is not constrained. Because Swan is confined to the upper half, the 72--66~km bin
as a whole shows only a weak signal ($5.9\pm1.3$\,W\,sr$^{-1}$ at $T_{\rm rot}=T_{\rm vib}=\SI{7500}{K}$). Below 66~km Swan is not detected,
while the CN $\Delta v=-1$ band decreases from 33 to \SI{8}{W\,sr^{-1}} (400--\SI{440}{nm}, 66--60 to 55--49~km); we do not quote upper limits there, because they depend
on the assumed band temperature and on the continuum polynomial, with which a weak band is degenerate (see the
72--69~km fit above).

\subsubsection{N$_2^+$ first negative system}
\label{sec:n2p}
N$_2^+$ B--X (0,0) at \SI{391.4}{nm} was searched for in the second-order spectra (\SIrange{383}{396}{nm}),
fitting each exposure with CN $\Delta v=0$ and N$_2^+$ (0,0)+(1,1) (details in Appendix~\ref{app:n2p}). CN alone does not
reproduce the red side of the \SIrange{388}{392}{nm} profile; adding N$_2^+$ halves $\chi^2$ at 66--55~km, and the
atomic alternatives (first-order O~I \SI{777}{nm}, Si~I \SI{390.55}{nm}) fit worse or are excluded by the
infrared-cut filter and the composition of the ablator. Combining 66--60 and 60--55~km gives
N$_2^+$(0,0)/CN($\Delta v=0$) $=0.25\pm0.05$ ($0.31\pm0.06$ for $T_{\rm N_2^+}=\SI{10000}{K}$); at 80--72~km
N$_2^+$ is not detected ($<0.42$; Figure~\ref{fig:n2p}).

\subsection{Atomic lines and electron density}
\label{sec:lines}
\begin{figure}[tbp]
\centering
\includegraphics[width=0.85\textwidth]{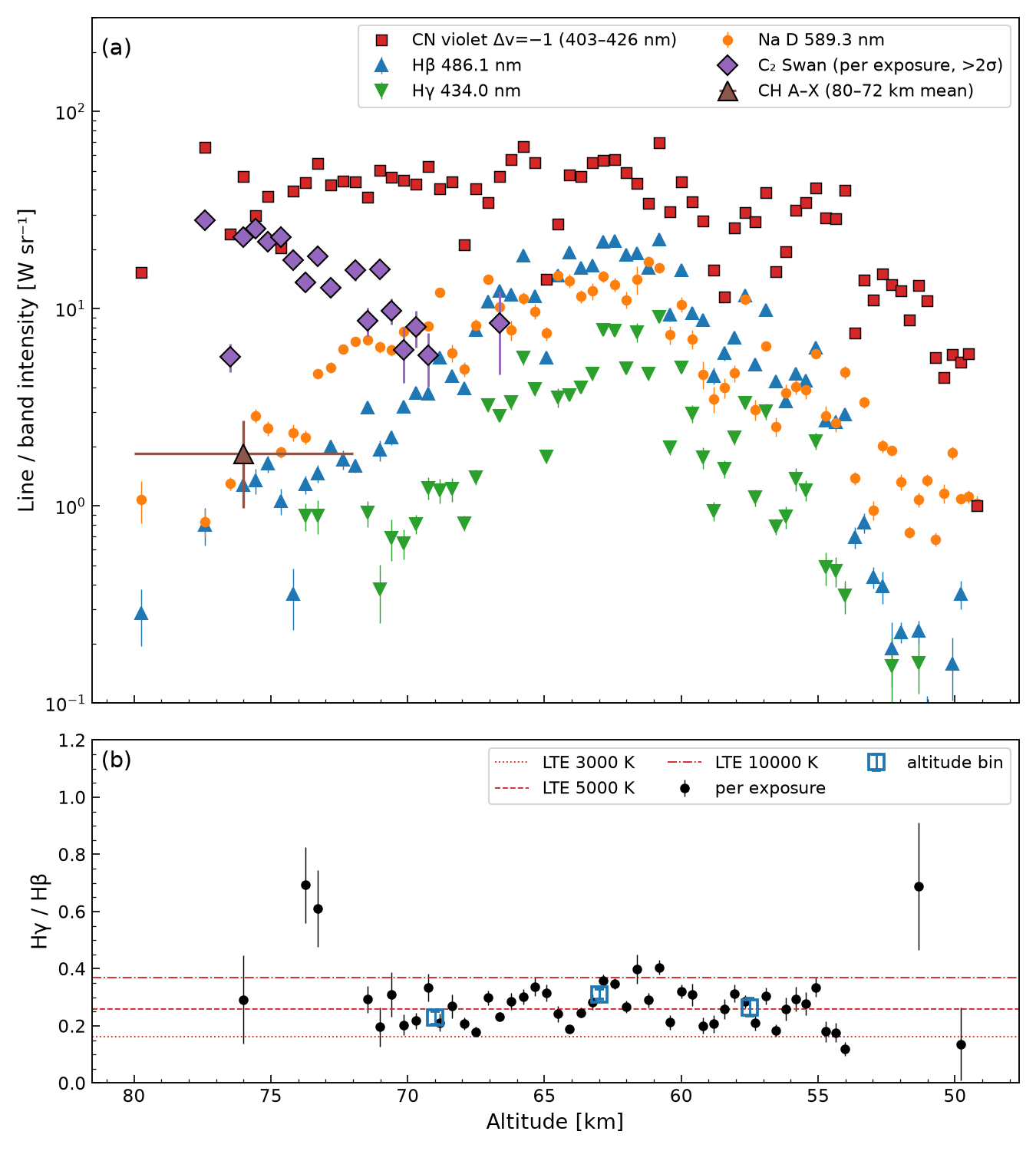}
\caption{\textbf{Intensities and equivalent widths of the atomic lines and the CN band.} (a) Radiant intensities of H$\beta$, H$\gamma$, Na~D, the CN $\Delta v=-1$ band, the C$_2$ Swan band
(per exposure where above 2$\sigma$, with the temperatures of the 80--72~km fit; Section~\ref{sec:swan}) and the CH A--X band (80--72~km mean; Section~\ref{sec:ch}). (b)
H$\gamma$/H$\beta$ per exposure and in altitude bins (squares), with optically thin LTE values for 3000, 5000
and \SI{10000}{K}.}
\label{fig:lines}
\end{figure}

Line intensities were measured in every exposure by fitting a Gaussian of the exposure's line-spread width
on a local linear continuum (H$\beta$, H$\gamma$, Na~D), and the CN $\Delta v=-1$ band by integrating above a
linear baseline (\SIrange{403}{426}{nm}) (Figure~\ref{fig:lines}). Because the measured line-spread function has
broad wings, a Gaussian of its width does not collect the whole line: for O~I \SI{615.7}{nm} (see below) the integral
with the measured line-spread function is 2.1--2.2 times the Gaussian intensity. The line intensities quoted in this
paper are therefore lower limits to the total line emission, by up to a factor of about 2, while ratios of lines
measured in the same way are not affected. H$\beta$ and H$\gamma$ reach their
maximum in the exposure of peak brightness (60.8~km; H$\beta$ \SI{22}{W\,sr^{-1}}), Na~D one exposure earlier (61.2~km,
\SI{17}{W\,sr^{-1}}; \SI{16}{W\,sr^{-1}} at peak brightness).
The equivalent width of H$\beta$ stays at about \SI{2}{nm} from 72 to 60~km and then decreases to
\SI{1.0}{nm} at 60--55~km and \SI{0.5}{nm} below 55~km, i.e.\ hydrogen emission follows the heating pulse and
declines faster than the continuum after peak heating. The CN band behaves differently: its intensity
scatters around \SI{44}{W\,sr^{-1}} (median; single exposures 14--\SI{66}{W\,sr^{-1}}) from 78 to 62~km, without the rise
of the continuum, declines below about 62~km (one exposure, at peak brightness, gives \SI{69}{W\,sr^{-1}}, but its neighbours give
31--\SI{34}{W\,sr^{-1}}), and its ratio to the continuum at \SI{550}{nm} falls from \SI{31}{nm} (80--72~km) to \SI{2.5}{nm}
(60--55~km). This integral (single exposures, \SIrange{403}{426}{nm} above a linear baseline) is a different
measure from the band intensity of the METEORAD fits (co-added exposures, \SIrange{400}{440}{nm};
Section~\ref{sec:cn}), which is used for the temperatures and the Swan/CN ratio; the two follow the same trend but
differ in absolute value. H$\alpha$ (\SI{656.3}{nm}) is also detected, beyond the infrared-cut filter where the relative
sensitivity is about 1.5\,\% of its maximum; because the sensitivity there is uncertain, we do not analyse its
intensity. Its equivalent width, which does not depend on the sensitivity, is comparable to that of H$\beta$
(1--\SI{3}{nm}) from 70 to 55~km.

The remaining features (Figure~\ref{fig:altspec}) were examined with METEORAD spectra of 22 air and 20 metal and
oxide species (details in Appendix~\ref{app:features}). O~I \SI{615.7}{nm} and \SI{533.0}{nm} are identified at 66--55~km;
a line at \SI{568.15}{nm} at 52--49~km is probably the Na~I 568.3/\SI{568.8}{nm} doublet, which would require a
strongly self-absorbed Na~D; and Ca~I \SI{422.7}{nm} is seen as a separate line below 55~km. The feature at
448--\SI{449}{nm} belongs to the excess in the CN $\Delta v=-2$ range, the residual at 585--\SI{587}{nm} is the wing
of Na~D, and a weak line at \SI{636.7}{nm} is not identified.

The ratio H$\gamma$/H$\beta$ (energy units) is 0.23, 0.31 and 0.27 ($\pm0.02$--0.03) in the bins 72--66,
66--60 and 60--55~km. For optically thin emission, the intensity of a line from the upper level $u$ is
$I_{ul}=(hc/4\pi\lambda_{ul})\,A_{ul}\,n_u\,L$, and with Boltzmann populations $n_u\propto g_u\exp(-E_u/kT_{\rm exc})$
(Appendix~\ref{app:temps}) the ratio of two lines gives the excitation temperature,
\begin{equation}
\begin{split}
\frac{I_{\rm H\gamma}}{I_{\rm H\beta}}&=\frac{g_5A_{52}\lambda_{42}}{g_4A_{42}\lambda_{52}}
\exp\!\left(-\frac{E_5-E_4}{kT_{\rm exc}}\right),\\
T_{\rm exc}&=\frac{E_5-E_4}{k\,\ln\!\left[(g_5A_{52}\lambda_{42})/(g_4A_{42}\lambda_{52})\big/(I_{\rm H\gamma}/I_{\rm H\beta})\right]},
\end{split}
\label{eq:texc}
\end{equation}
with $g_5=50$, $g_4=32$, $A_{52}=2.530\times10^{6}$\,s$^{-1}$, $A_{42}=8.419\times10^{6}$\,s$^{-1}$, $E_5-E_4=\SI{0.306}{eV}$
(NIST). The measured ratios correspond to $T_{\rm exc}=4300$, 6700 and \SI{5200}{K} (range \SIrange{3800}{7500}{K}); the
per-exposure values lie between the curves for 3000 and \SI{10000}{K} (Figure~\ref{fig:lines}).

The H$\beta$ line is broader than the line-spread function derived from the zero-order image, but so is
Na~D, whose Stark width is negligible: the difference between the point-spread functions at the zero-order and
the dispersed positions adds an instrumental width. We therefore calibrate the instrumental profile of each
exposure on Na~D and fit H$\beta$ with the same profile. The instrumental profile is the measured line-spread
function convolved with a Gaussian of width $\sigma_x$, which is fitted on Na~D modelled as the doublet 588.995
and \SI{589.592}{nm} (intensity ratio free between 1 and 2, common shift, linear continuum). H$\beta$ is then
fitted with this profile convolved with a Lorentzian of FWHM $\Delta\lambda_{1/2}$, the Stark width, which is
converted with the H$\beta$ Stark relation of computer-simulated profiles \citep{Gigosos2003},
\begin{equation}
N_e=10^{23}\left(\frac{\Delta\lambda_{1/2}}{4.8\,{\rm nm}}\right)^{1/0.68}\ {\rm m}^{-3}.
\label{eq:stark}
\end{equation}
The 23 exposures with H$\beta$ and Na~D above 10$\sigma$ (73--56~km) give $\sigma_x=0.005$--\SI{2.2}{nm}, and with
this profile the H$\beta$ width is fully explained by the instrument: the summed $\chi^2$ profiles of the
altitude bins have their minima at $\Delta\lambda_{1/2}=0$--\SI{0.2}{nm} and 2$\sigma$ upper limits of 0.31, 0.02
and \SI{0.045}{nm} at 72--66, 66--60 and 60--55~km, i.e.\ $N_e<2\times10^{21}$, $3\times10^{19}$ and
$1\times10^{20}$\,m$^{-3}$. The limits do not depend on the Na~D intensity ratio, change little when $\sigma_x$ is
scaled with wavelength ($<0.08$\,nm at 66--55~km) or when a Gaussian instrumental profile is used, and are the
weakest for a Gaussian instead of a Lorentzian Stark profile (0.985, 0.34 and \SI{0.375}{nm}; $N_e<1\times10^{22}$\,m$^{-3}$ at
72--66~km and $<2\times10^{21}$\,m$^{-3}$ at 66--55~km), which we adopt as the conservative limit.

For comparison, the electron density of air in ionization equilibrium follows from the Saha equation,
\begin{equation}
\frac{n_e\,n_{i}}{n_{0}}=\frac{2U_{i}(T)}{U_{0}(T)}\left(\frac{2\pi m_ekT}{h^2}\right)^{3/2}
\exp\!\left(-\frac{\chi}{kT}\right),
\label{eq:saha}
\end{equation}
with ground-term partition functions $U$ (N 4, N$^+$ 9, O 9, O$^+$ 4), the ionization energies $\chi$ of N (\SI{14.53}{eV})
and O (\SI{13.62}{eV}), atomic fractions of 0.79 (N) and 0.21 (O) and charge neutrality.
For the post-shock density at 60~km (about $10^{23}$\,m$^{-3}$ atoms, 15 times the ambient density of
NRLMSISE-00), Eq.~\ref{eq:saha} gives $N_e=10^{18}$, $1.5\times10^{20}$, $7\times10^{20}$ and $7\times10^{21}$\,m$^{-3}$ at
5000, 7000, 8000 and \SI{10000}{K}. The conservative limit $N_e<2\times10^{21}$\,m$^{-3}$ for the gas that emits
H$\beta$ therefore corresponds to an equilibrium temperature below about \SI{8800}{K}, consistent with the
H$\gamma$/H$\beta$ excitation temperature and with emission from the boundary layer rather than from the hottest
part of the shock layer (Section~\ref{sec:hydrogen}).

\subsection{Surface temperature: colour and brightness temperatures}
\label{sec:continuum}
\begin{figure}[tbp]
\centering
\includegraphics[width=\textwidth]{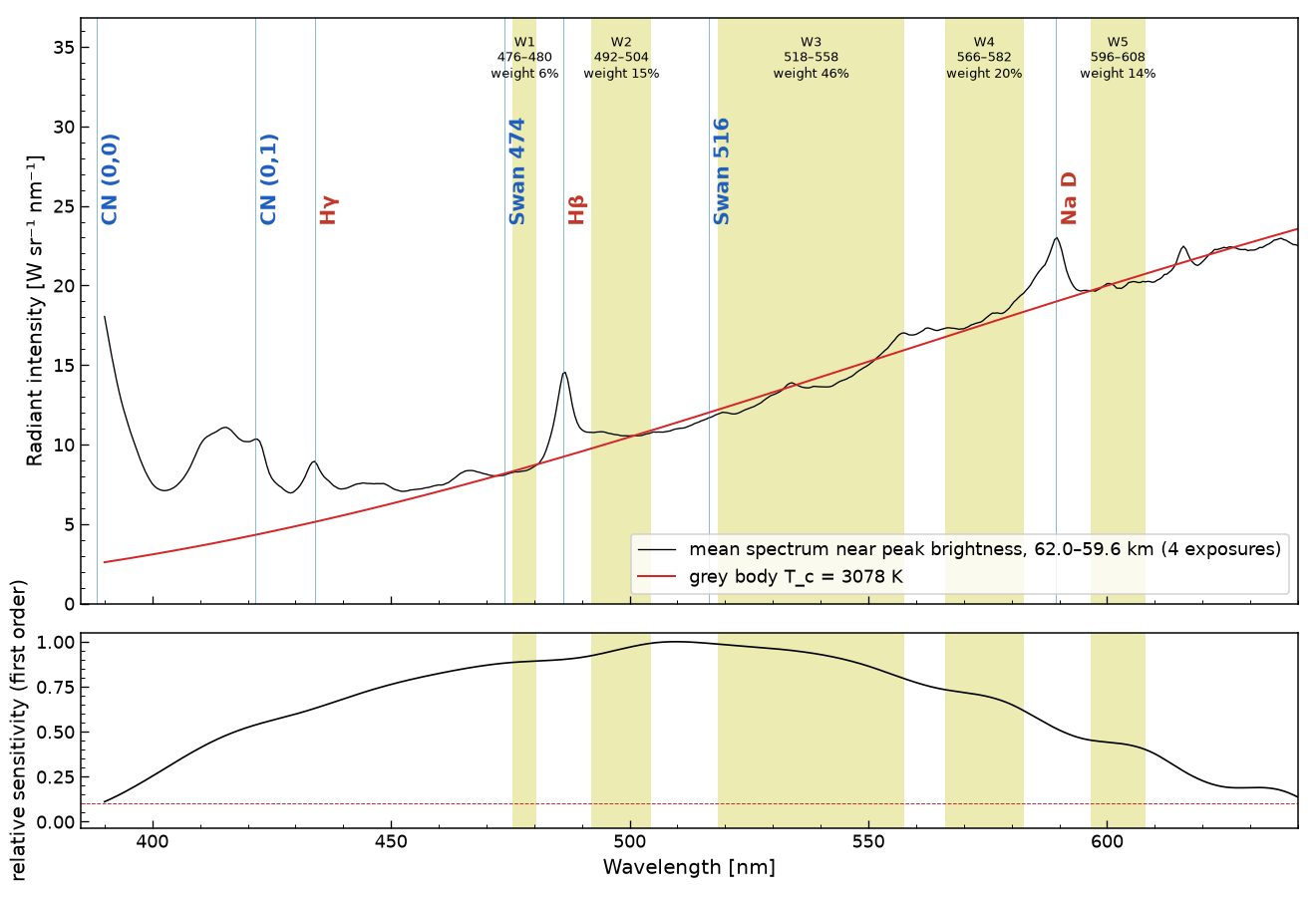}
\caption{\textbf{Continuum fit windows on the spectrum at peak brightness.} Fit windows (yellow) on the mean spectrum of four exposures within \SI{1}{s} of peak brightness
(62.0, 60.8, 60.0 and 59.6~km, $t=11.4$--13.0~s), selected for a continuum signal-to-noise ratio of at least
55, a reduced $\chi^2$ below 0.6, a line-spread function of at most \SI{3.1}{nm} and a wavelength correction from
three lines (Appendix~\ref{sec:wavecorr}),
with the grey-body fit (red). Blue lines: CN (0,0) and (0,1) heads,
H$\gamma$, C$_2$ Swan, H$\beta$ and Na~D. Bottom: relative first-order sensitivity.}
\label{fig:windows}
\end{figure}

The continuum of each exposure is fitted with a grey body,
\begin{equation}
I(\lambda) = \varepsilon A\,B(\lambda,T),
\label{eq:model}
\end{equation}
where $B$ is the Planck function and $\varepsilon A$ the effective emitting area. The fit uses five windows inside
the range where the relative sensitivity exceeds 40\,\% (\SIrange{409.5}{608}{nm}) and that avoid all
emission lines and bands: 475.5--480.5, 492--504.5, 518.5--557.5, 566--582.5 and 596.5--608\,nm
(Figure~\ref{fig:windows}). The range 439--\SI{466}{nm} is excluded because it contains the CN B--X $\Delta v=-2$
sequence (heads at 460.6, 457.8 and \SI{455.3}{nm}, degraded to the violet; Appendix~\ref{sec:shape}).
Weights include the noise and a 5\,\% model term; statistical errors come from a wild block bootstrap of the
residuals, and systematic errors from a jackknife of the Sirius exposures (\SI{81}{K} at peak brightness),
the aerosol optical depth (0.03 to 0.15, \SI{2}{K}), the choice of windows (\SI{40}{K}) and stray light
(\SI{60}{K}, Appendix~\ref{sec:stray}). The spectra are corrected as described in Appendix~\ref{sec:selfcal}.
Exposures with a continuum signal-to-noise ratio of at least 20 are used: 47 exposures from 73.7 to 50.7~km.

\begin{figure}[tbp]
\centering
\includegraphics[width=\textwidth]{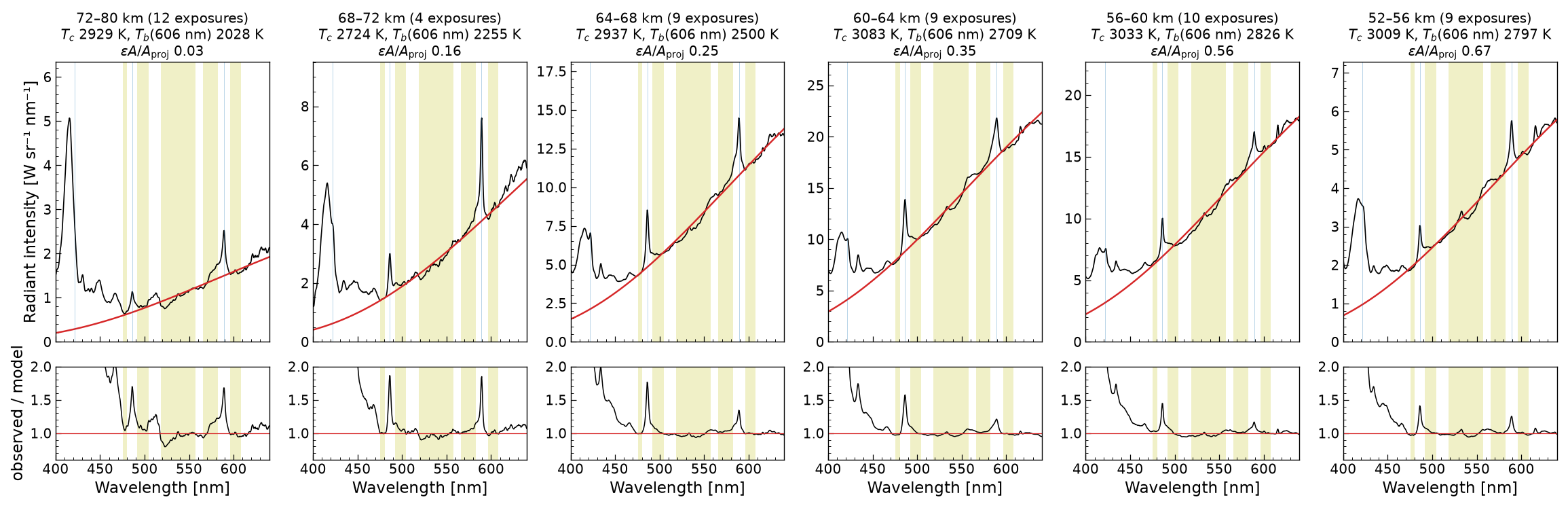}
\caption{\textbf{Grey-body fits to the mean spectra in altitude bins.} Mean spectra in altitude bins with the grey-body fit (red). Bottom: ratio of observed to model
spectrum. Yellow: fit windows. The 80--72~km bin, above the
range of the temperature history, co-adds all its 12 exposures; its fit ($T_c=\SI{2930}{K}$, reduced $\chi^2=3.7$)
is affected by the C$_2$ Swan bands in the windows (Section~\ref{sec:swan}).}
\label{fig:bgfits}
\end{figure}

The grey body fits the windows with a median reduced $\chi^2$ of 0.6 (Figure~\ref{fig:bgfits}). Leaving out any
one window changes the colour temperature by $-60$ to $+40$\,K (medians over the exposures); the alternative
window sets tested give higher values, \SIrange{3250}{3430}{K} at peak brightness, because all of them include part
of the CN $\Delta v=-2$ region or the wings of H$\gamma$ and Na~D. A pedestal constant in detector counts, bounded by the counts at
the ends of the trace, changes the colour temperature at peak brightness by less than \SI{2}{K}; without the correction of
Appendix~\ref{sec:selfcal} the colour temperature at peak brightness is \SI{2968}{K}.

Two temperatures are derived from the continuum, using complementary information.
The colour temperature $T_c$ is the temperature $T$ of the fit of Eq.~(\ref{eq:model}): it is set by the
spectral shape, i.e. by the relative intensities in the windows, and does not depend on the absolute
calibration, the range, the emitting area or a wavelength-independent emissivity; it does depend on the
shape of the sensitivity curve. The brightness temperature $T_b$ is the
temperature of a black body (emissivity one) that would produce the observed absolute radiance in a band,
\begin{equation}
B(\lambda_c, T_b) = \frac{\langle I(\lambda)\rangle_{\rm band}}{A_p},
\label{eq:tb}
\end{equation}
where the radiant intensity is averaged over the band and $\lambda_c$ is the band centre.
$A_p$ is the projected area of the visible forebody, a \SI{45}{\degree} sphere-cone with a base diameter of
\SI{0.413}{m} and a nose radius of \SI{0.207}{m} \citep{Prabhu2022}, seen at the angle $\theta$ between the
velocity and the line of sight (\SI{19}{\degree} at 80~km to \SI{134}{\degree} at 49~km). Four bands are used,
445--480, 540--555, 600--612 and 622--\SI{640}{nm} (the band at 445--\SI{480}{nm} contains the weak CN
$\Delta v=-2$ and C$_2$ Swan $\Delta v=+1$ bands, which are kept out of the fit windows, so that its $T_b$ is an
upper limit); the band at 600--\SI{612}{nm} (``$T_b(606)$'') is quoted as
representative, since it lies in the last fit window and is free of emission. $T_b$ depends on the absolute
calibration, the range, the extinction and the transfer curve, but only weakly: at \SI{606}{nm} and
\SI{2800}{K}, $\Delta T_b/T_b = (\lambda T/c_2)\,\Delta L/L$, i.e. 12\,\% of the relative intensity error
(\SI{20}{K} for the 7\,\% calibration uncertainty). For a grey body with emissivity $\varepsilon<1$, or an emitting
area smaller than $A_p$, the surface temperature is higher than $T_b$, so $T_b$ is a lower limit if the forebody
is the only emitter. The two are linked by the fitted amplitude: with $T=T_c$, the effective emitting area
$\varepsilon A$ follows from Eq.~(\ref{eq:model}), and $\varepsilon A/A_p<1$ corresponds to $T_c>T_b$. The
agreement of $T_b$ between the bands within the window span tests the assumed grey-body shape independently of the fit.

\begin{table}[t]
\centering
\caption{\textbf{Continuum temperatures in altitude bins.} (S-Log3; grey body, windows of
Figure~\ref{fig:windows}). $T_c$: colour temperature with the error of the mean; $\sigma_{\rm sys}$:
systematic error; $T_b$: brightness temperature at \SI{606}{nm}; $\varepsilon A/A_p$: emitting area over the
projected forebody area; last column: $T_c$ with the S-Log2 and Rec.~709 transfer curves.}
\label{tab:temp}
\resizebox{\textwidth}{!}{%
\begin{tabular}{lcccccccc}
\toprule
Altitude (km) & $t$ (s) & $\theta$ (\si{\degree}) & $n$ & $T_c$ (K) & $\sigma_{\rm sys}$ (K) & $T_b$ (K) & $\varepsilon A/A_p$ & $T_c$ S-Log2/Rec.709 (K)\\
\midrule
72--68 & 5.5--6.6 & 28--31 & 4 & $2720\pm10$ & 110 & 2260 & 0.16 & 2950/3010\\
68--64 & 7.4--10.0 & 33--44 & 9 & $2900\pm40$ & 100 & 2500 & 0.27 & 3170/3330\\
64--60 & 10.2--12.8 & 46--64 & 9 & $3080\pm10$ & 110 & 2710 & 0.36 & 3400/3580\\
60--56 & 13.0--15.6 & 66--91 & 10 & $3010\pm20$ & 100 & 2830 & 0.57 & 3270/3410\\
56--52 & 15.8--18.6 & 94--118 & 9 & $3010\pm20$ & 100 & 2800 & 0.54 & 3210/3270\\
\bottomrule
\end{tabular}}
\end{table}

The results are listed in Table~\ref{tab:temp} and shown in Figure~\ref{fig:history}. At peak brightness
the colour temperature is $T_c=\SI{3042 \pm 61}{K}$ (statistical) $\pm\SI{108}{K}$ (systematic), and the
brightness temperatures in the four bands from 445 to \SI{640}{nm} agree within \SI{90}{K} (2853, 2801, 2774 and
\SI{2765}{K} at 445--480, 540--555, 600--612 and 622--\SI{640}{nm}). A weighted linear fit over the 47 exposures
gives mean rates of $\SI{15 \pm 4}{K\,s^{-1}}$ for the colour temperature and $\SI{42 \pm 4}{K\,s^{-1}}$ for the
brightness temperature. The rise of the colour temperature is concentrated above 60~km: from about \SI{2720}{K} at 70~km to \SI{3080}{K} at 64--60~km, after which it
stays at about \SI{3010}{K}; the brightness temperature rises from 2260 to \SI{2830}{K},
so that $T_c-T_b$ decreases from 470 to \SI{190}{K}. The emitting area is 0.06 to 0.74 (median 0.41; two exposures
at 51.0 and 50.7~km, with reduced $\chi^2$ of 6--10, give 1.7 and 2.6) of the projected forebody area and increases
from 0.16 at 72--68~km to 0.57 at 60--56~km. The S-Log2 and Rec.~709 transfer curves give colour temperatures
higher by 200--\SI{320}{K} and 260--\SI{500}{K}, respectively.

\subsection{Luminous efficiency and heating}
\label{sec:lum}
\begin{figure}[tbp]
\centering
\includegraphics[width=\textwidth]{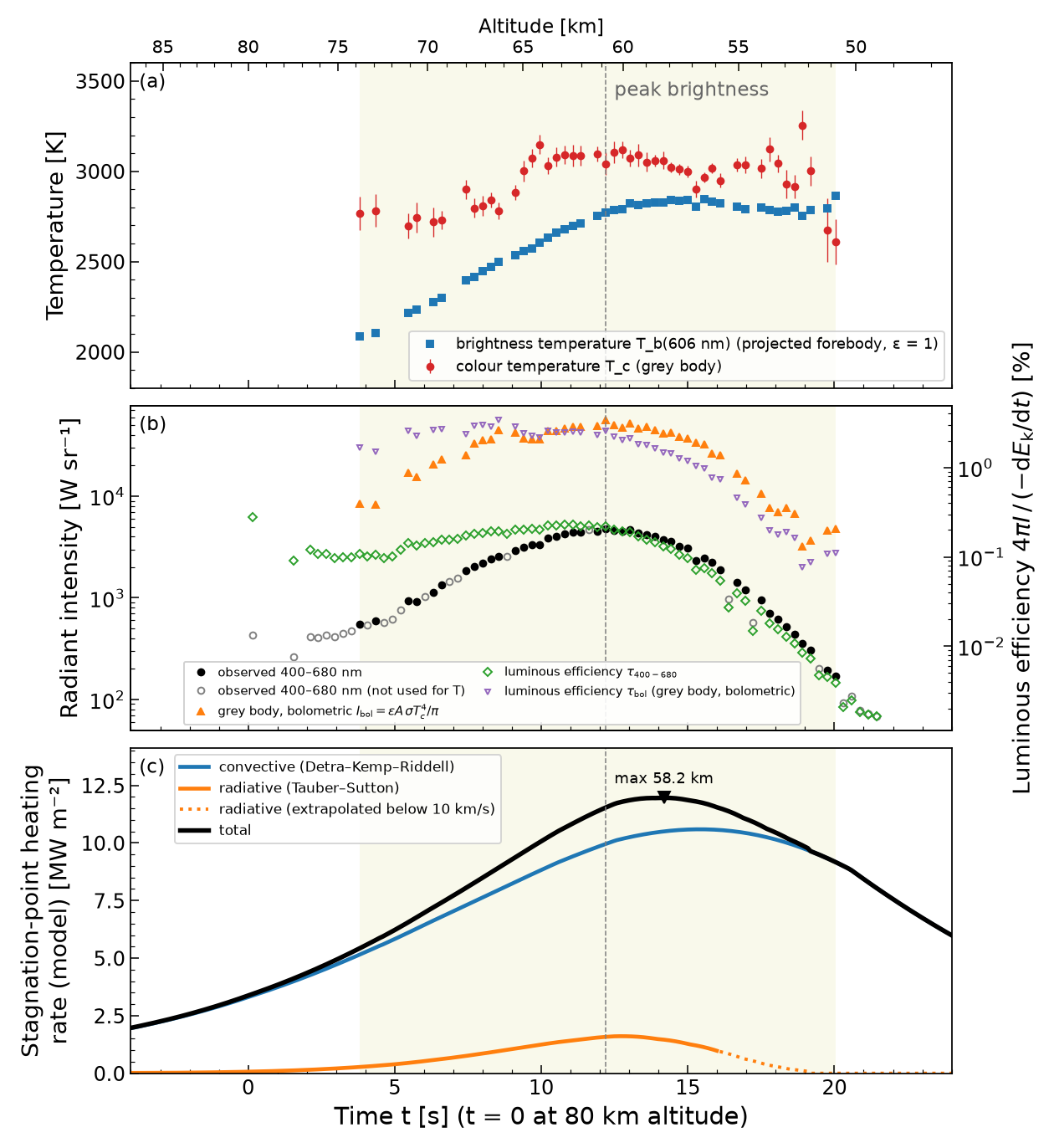}
\caption{\textbf{Temperatures, radiant intensity, luminous efficiency and modelled heating rate versus time.} Time history (time $t$ after the capsule passed 80~km; top axis: altitude). (a) Colour temperature $T_c$ (grey body) and brightness temperature
$T_b$ at \SI{606}{nm}. (b) Observed radiant intensity at 400--\SI{680}{nm} (filled: exposures used for the
temperatures) and bolometric radiant intensity of the fitted grey body, $I_{\rm bol}=\varepsilon A\,\sigma T_c^4/\pi$ (left
axis), and the luminous efficiencies $\tau_{400-680}$ and $\tau_{\rm bol}$ of Eq.~\ref{eq:tau} for $m=\SI{16}{kg}$
(right axis). (c) Modelled stagnation-point heating rate along the pre-flight trajectory (Detra--Kemp--Riddell convective, Tauber--Sutton radiative, dotted where extrapolated below
\SI{10}{km\,s^{-1}}; $R_n=\SI{0.207}{m}$, NRLMSISE-00; Appendix~\ref{app:heating}). Dashed line:
peak brightness; shaded: exposures used for the temperatures.}
\label{fig:heating}
\end{figure}

Figure~\ref{fig:heating} places the temperatures and the emission against a simple model of the stagnation-point
heating along the pre-flight trajectory: the Detra--Kemp--Riddell convective correlation
\citep{DetraKempRiddell1957} and the Tauber--Sutton radiative correlation \citep{TauberSutton1991} for a nose radius
of \SI{0.207}{m}, with the density of NRLMSISE-00 \citep{Picone2002} (details in Appendix~\ref{app:heating}).
The radiative relation is stated for 10--\SI{16}{km\,s^{-1}} and nose radii of 0.3--\SI{3}{m} and is extrapolated here.
The modelled heating rate peaks at \SI{12.0}{MW\,m^{-2}} at 58.2~km ($t=14.2$~s, 17:28:52.6~UTC, \SI{10.4}{km\,s^{-1}};
convective \SI{10.6}{MW\,m^{-2}} at 56.7~km, radiative \SI{1.6}{MW\,m^{-2}} at 60.1~km), \SI{2}{s} after the
peak of the observed radiant intensity. This agrees with the pre-flight prediction of the stagnation-point
heat flux of the capsule, which peaks at \SI{12.2}{MW\,m^{-2}} at 72.9~s after the entry interface at 200~km, i.e.\ at
58.5~km on the pre-flight trajectory \citep{Yamada2021}. The Sutton--Graves convective correlation \citep{SuttonGraves1971} gives
15\,\% less at nearly the same altitude. For the Hayabusa capsule (same shape and size, \SI{12.0}{km\,s^{-1}}), the
pre-flight predictions were about 16 and \SI{4}{MW\,m^{-2}} (cold-wall convective and radiative), reduced slightly by
radiation coupling in the flow-field reconstruction \citep{Fujita2011AIAA}. The colour temperature
reaches its plateau before the heating peak and does not follow its decline, as expected from the thermal
inertia of the charring ablator. The brightness temperature stays near \SI{2800}{K} after the peak, and the
observed intensity falls mainly because the forebody turns away from the observer (aspect angle beyond
\SI{90}{\degree}), which reduces its projected area. We define two luminous efficiencies relative to the loss rate of the kinetic energy $E_{\rm k}=mV^2/2$ of the
capsule, one for the observed band and one for the bolometric emission of the fitted grey body:
\begin{equation}
\begin{split}
\tau_{400-680}&=\frac{4\pi\int_{400\,{\rm nm}}^{680\,{\rm nm}}I_\lambda\,{\rm d}\lambda}{-{\rm d}E_{\rm k}/{\rm d}t},\\
\tau_{\rm bol}&=\frac{4\pi I_{\rm bol}}{-{\rm d}E_{\rm k}/{\rm d}t},\\
I_{\rm bol}&=\varepsilon A\int_0^\infty B(\lambda,T_c)\,{\rm d}\lambda=\frac{\varepsilon A\,\sigma T_c^4}{\pi},
\end{split}
\label{eq:tau}
\end{equation}
with $-{\rm d}E_{\rm k}/{\rm d}t=-mV\,{\rm d}V/{\rm d}t$ from the trajectory, $m=\SI{16}{kg}$, $I_\lambda$ the
observed radiant intensity, $\varepsilon A$ and $T_c$ from the grey-body fit (Eq.~\ref{eq:model}) and $\sigma$ the
Stefan--Boltzmann constant; $4\pi I$ is a nominal radiated power for isotropic emission. $I_{\rm bol}$ extrapolates
the grey body to the infrared and is about twelve times the observed band, since most of the thermal emission of
a \SI{3000}{K} surface lies in the infrared. With the maximum deceleration of \SI{42}{g} at 50~km,
$\tau_{400-680}$ is about 0.2\,\% at peak brightness and $\tau_{\rm bol}$ 0.1--3\,\% (median 1.8\,\%); most of the
kinetic energy goes into the shock layer and the wake rather than into the surface emission. Integrated over 80--49~km,
the radiated energy in 400--\SI{680}{nm} is \SI{4.8e5}{J}, 0.09\,\% of the kinetic energy lost over the same interval
(\SI{11.6}{} to \SI{8.1}{km\,s^{-1}}, \SI{5.6e8}{J}) and 0.04\,\% of the kinetic energy at entry (\SI{1.09e9}{J} at 100~km).

\subsection{Wake and persistent train}
\label{sec:wake}
In meteors, a large part of the light can come from the wake behind the body, and bright meteors leave a
persistent train along their path \citep{Abe2009LNP}. We distinguish a wake that moves with the capsule from a train that stays fixed
in the air and fades after the capsule has passed.

\paragraph{Co-moving wake} The apparent direction of motion in the images was derived from consecutive plate
solutions; it is nearly constant during the observation (towards the upper left, Figure~\ref{fig:frame}), so a wake
would extend behind the capsule to the lower right, at about \SI{25}{m} per pixel at the range of peak brightness
(\SI{66}{km}; 21--\SI{68}{m} over the ranges of 57--\SI{180}{km}).
In the zero order (white light), the brightness 40--\SI{200}{px} behind the capsule (1--\SI{5}{km}) does not differ
from that ahead of or beside it, at a level of about 0.1\,\% of the capsule per pixel. Line emission of a wake
would appear in the first order as streaks displaced from the capsule trace, perpendicular to the dispersion, at
the line wavelengths. Images of Na~D minus the scaled continuum, aligned on the capsule's Na~D and rotated to the
wake direction, were stacked in altitude bins: no streak is seen, and 100--\SI{250}{px} behind the capsule
(2.5--\SI{6.5}{km}) the Na~D emission of a wake is below 6\,\% of the capsule's Na~D at 72--60~km and below 33\,\%
at 60--49~km (2$\sigma$). Within about \SI{2}{km} the wake cannot be separated from the capsule image, which is
smeared by the motion of the camera.

\begin{figure}[tbp]
\centering
\includegraphics[width=\textwidth]{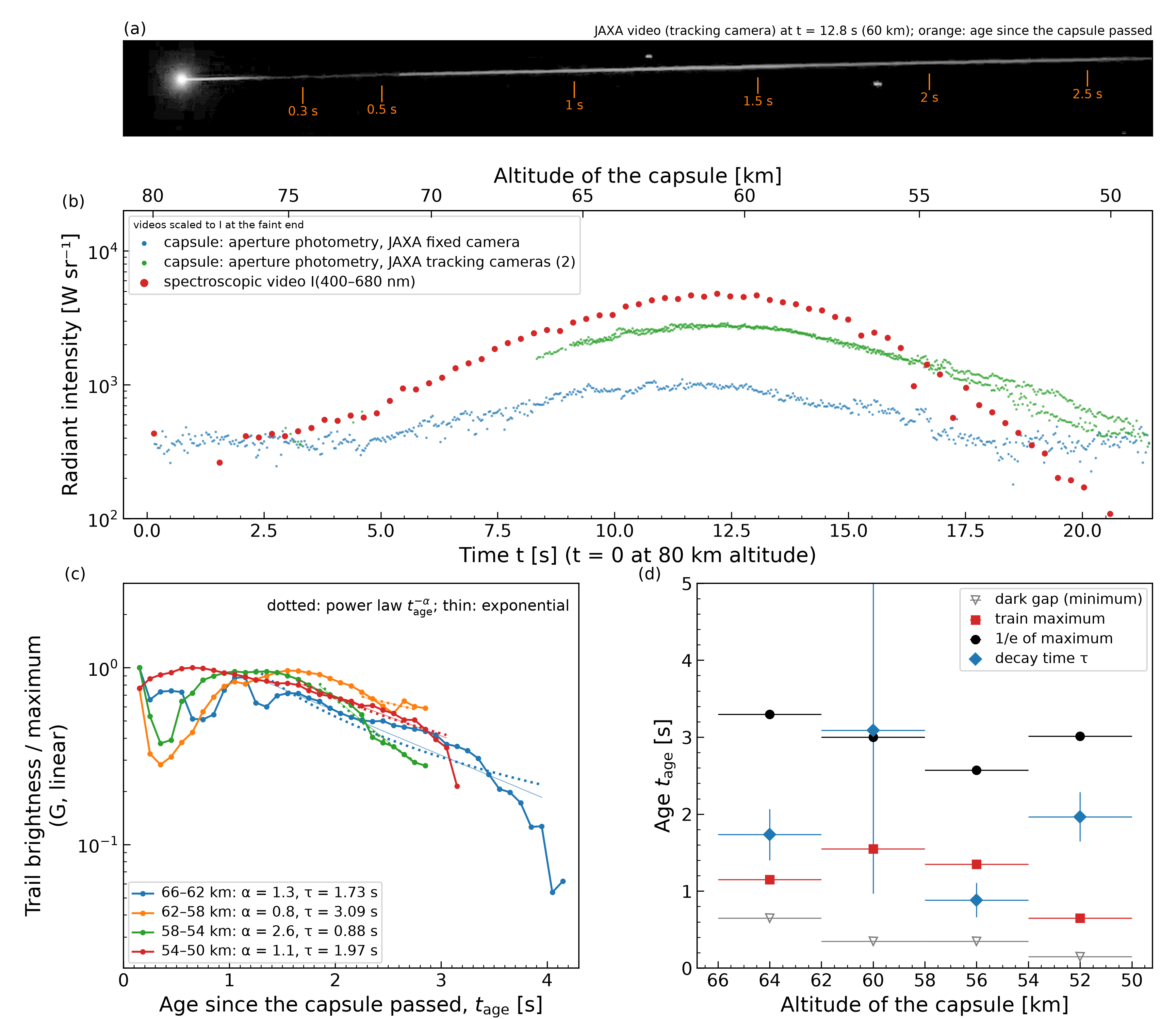}
\caption{\textbf{Persistent train of the capsule in the JAXA videos.} Persistent train of the capsule in the JAXA videos (\citealp{JAXAfireball2020}, credit: JAXA; the three videos are
combined). (a) Frame of a tracking camera at $t=12.8$~s (capsule near 60~km), with the age of the trail
(time since the capsule passed) derived from the drift of the stars. (b) Aperture photometry of the capsule in the
videos ($r\le15$~px, G, linearised with $\gamma=0.9$ from the stars; blue: fixed camera, green: the two tracking cameras), each scaled to the spectroscopic light curve $I_{400-680}$
(red) at its faintest frames: the heads are saturated and fall short of $I$ near the peak (0.2--0.6 of $I$), and below
55~km (after $t=18.6$~s) the trail merges with the head; top axis: altitude of the capsule. (c) Trail brightness (G,
linearised, corrected for vignetting, normalised to its maximum) against age, the videos averaged per altitude bin
of the capsule (logarithmic brightness, linear age); dotted: power-law fits $t_{\rm age}^{-\alpha}$, thin lines: exponential fits, both from 80
to 10\,\% of the maximum or the frame edge.
(d) Characteristic ages against altitude: the dark gap behind the head, the train maximum, the age at which the train
has faded to $1/e$ of its maximum (at 62--58~km the frame edge, a lower limit) and the decay time $\tau$ (error: fit and choice of
the fit window); in all bins the train is analysed behind the dark gap.}
\label{fig:train}
\end{figure}

\paragraph{Persistent train} A train is, however, seen in an independent video of the entry recorded by JAXA near
Coober Pedy with a fixed, unfiltered camera and published by the JAXA Hayabusa2 Project (video 1 of 3; credit: JAXA)
\citep{JAXAfireball2020} (1920$\times$1080 pixels, 29.97 frames per second).
The video was timed by matching the light curve of the capsule to our spectroscopic light curve
($\pm$\SI{0.2}{s}) and calibrated on 115 stars, which show that it is close to linear ($\gamma=0.9$); the head of
the capsule is saturated, so its brightness is taken from the spectroscopic light curve (details in Appendix~\ref{app:jaxa}).
 The capsule crosses the field
at 72--53~km. Along the path behind the capsule the sky becomes brighter than before the passage, i.e.\ the luminous
trail is fixed in the air, not attached to the capsule; in tracking videos of the same campaign the structure of the
trail moves with the stars. Because the camera is fixed, every point of the path gives the light curve of the train as
a function of its age $t_{\rm age}$, the time since the capsule passed. Stacked by the altitude of the passage
(Figure~\ref{fig:train}c), the train does not start at the capsule: it appears 0.5--\SI{0.9}{s} after the passage
(20\,\% of the maximum) and peaks at 1.1--\SI{1.9}{s}, for any tone curve between $\gamma=0.9$ and 2.7; the onset
follows the passage along the path and is not tied to the key frames of the compression, so the train is not an
artefact of the encoding. Its colour cannot be measured: the only colour reference in the frames, the capsule,
is saturated white (R:G:B = 1.01:1:1.01), and the 4:2:0 chroma subsampling of the 2-Mbit\,s$^{-1}$ video blends the
colour of a 1--2-px line with its surroundings. The train is barely resolved: its Gaussian width,
$\sigma=1.3$--1.9~px at ages of 1--\SI{2}{s} (up to 2.5~px at 72--66~km), exceeds that of the stellar images (0.95~px),
which leaves $\sigma\approx0.9$--1.6~px, i.e.\ about 85--\SI{155}{m} (1~px $\approx$ \SI{95}{m} at \SI{70}{km}).

\paragraph{Length of the train in the tracking videos} The other two JAXA videos \citep{JAXAfireball2020} were taken
with cameras that followed the capsule by hand; the train appears as a line behind the head
(Figure~\ref{fig:train}a). Its length converts to age without a plate solution, because the star field drifts
through the frame at the angular rate of the capsule relative to the camera (200--\SI{800}{px\,s^{-1}}, from the
frame-to-frame shift of the stars): with the cumulative star shift $\mathbf{S}(t)$ and the head position
$\mathbf{h}(t)$, the path of the head in sky-fixed pixels, $\mathbf{h}-\mathbf{S}$, gives the time at which the capsule
passed every point of the trail in a frame, and hence its age. The videos were timed as the static one (to about
$\pm$\SI{0.5}{s}, the heads being saturated); the ages do not depend on this timing. The brightness along the trail
was linearised ($\gamma=0.9$, assumed to hold for all three videos, which went through the same publication encoding)
and corrected for the vignetting of the lenses, measured with stars carried across the
frame by the pan (the throughput falls to 0.6--0.8 at the frame edges); profiles stop \SI{0.1}{s} before the edge,
which the trail reaches at ages of 1.2--\SI{4}{s}. The light curves of the three videos, normalised to their
maxima, were then averaged per altitude bin of the capsule (Figure~\ref{fig:train}c,d). In every bin the train is
analysed behind the dark gap that follows the head, whether the gap is deep or not: the maximum is taken after the
minimum of the gap, and the decay time was
fitted to the logarithm of the mean curve from 80 to 10\,\% of its maximum, or to the frame edge. At 62--54~km the
train is separated from the head by a dark gap at \SI{0.35}{s} (30--40\,\% of the train maximum), is brightest at
1.35--\SI{1.55}{s} and decays with $\tau=0.9\pm0.1$~s at 58--54~km and 2--3~s at 62--58~km (poorly constrained:
there the curve falls below 80\,\% of its maximum only at \SI{2.2}{s}), and fades to $1/e$ of its maximum at
\SI{2.6}{s} at 58--54~km, while at 62--58~km it is still above $1/e$ at the frame edge (\SI{3.0}{s}). At 66--62~km the gap is shallower (0.6 of the maximum at \SI{0.65}{s}), the maximum comes at
\SI{1.15}{s} and $\tau=1.7\pm0.2$~s ($1/e$ at \SI{3.3}{s}); at 54--50~km the gap is filled in (0.8 of the maximum),
the maximum moves to \SI{0.65}{s} and $\tau=2.0\pm0.2$~s ($1/e$ at \SI{3.0}{s}). The decay time therefore lies between 0.9 and about 3~s,
is shortest at 58--54~km and shows no trend with altitude (Section~\ref{sec:traindisc}); with $\gamma=1$ the values
change by less than 10\,\%. On a log--log scale the decline is not a straight line either: power laws
$I\propto t_{\rm age}^{-\alpha}$ over the same window give $\alpha=2.6\pm0.2$ at 58--54~km and 0.8--1.3 in the
other bins, and fit worse than or as well as the exponential (rms of $\ln I$ 0.066 against 0.060 at 58--54~km, 0.07--0.25
against 0.06--0.19 in the other bins). Over the short range of ages available (a factor of 2--3) the decay is therefore
best described by an e-folding time, with no evidence for a power-law tail. The visible train therefore spans
more than \SI{3}{s} of the passage, i.e.\ more than 30~km of the path (4--6~km in altitude) behind the
capsule, of which the first 3--4~km are dark at 62--54~km.

The light of the capsule therefore comes from the forebody shock layer, the boundary layer and the
surface; there is no detectable co-moving wake, and the air along the path glows only faintly, as a persistent
train that lights up about \SI{1}{s} after the passage, fades with $\tau\approx1$--\SI{3}{s}, and is visible over more
than 30~km of the path. Its origin is discussed in Section~\ref{sec:traindisc}.

\section{Discussion}
\label{sec:discussion}

\begin{figure}[tbp]
\centering
\includegraphics[width=\textwidth]{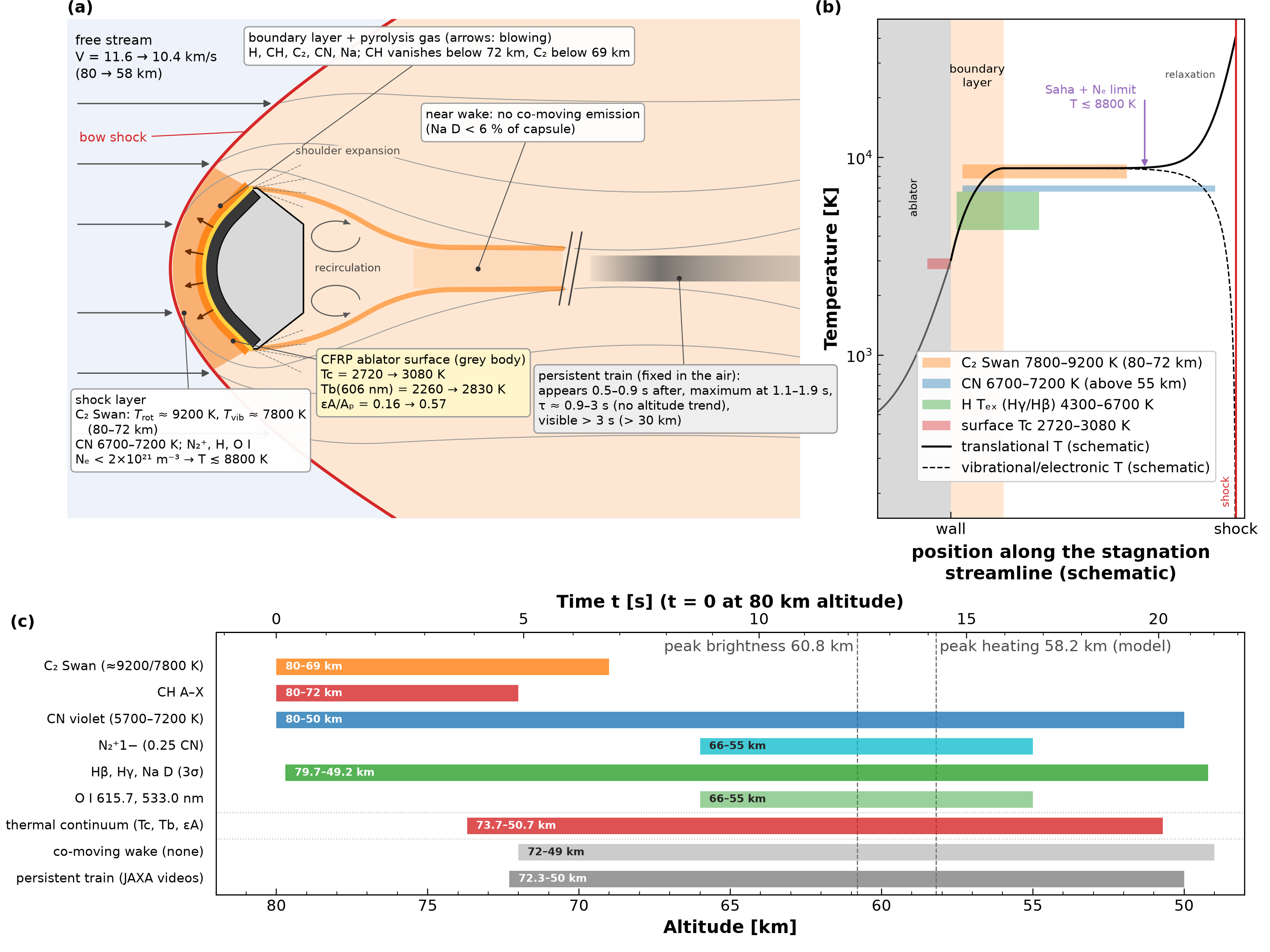}
\caption{\textbf{Flow around the capsule and summary of the measured quantities.} Schematic of the flow around the Hayabusa2 capsule and the results of this work (not to scale; the
shock standoff, about 0.07\,$R_n$, and the boundary layer are exaggerated). (a) Capsule frame with the flow from the
left: bow shock, shock layer, boundary layer with the pyrolysis gas of the CFRP ablator, shoulder
expansion, near-wake recirculation, the near wake without co-moving emission and, beyond a break (wake shortened),
the persistent train fixed in the air, annotated with the measured temperatures and species. (b) Qualitative
temperature along the stagnation streamline (translational and vibrational/electronic temperatures, shape only) with
the measured temperatures as bands: C$_2$ Swan (80--72~km), CN, the Balmer excitation temperature and the colour
temperature of the surface; the arrow marks the upper bound from the H$\beta$ Stark width and the Saha equation (the profiles are given in Appendix~\ref{app:schematic}). (c)
Altitude ranges of the detections (top axis: time $t$ after the capsule passed 80~km); dashed lines mark the peak
brightness (60.8~km) and the modelled peak heating (58.2~km).}
\label{fig:schematic}
\end{figure}

Figure~\ref{fig:schematic} summarises where in the flow around the capsule the measured quantities arise and over
which altitudes they were detected; the discussion below follows the same order, from the shock and boundary layers to the heat-shield
surface, the energy budget and the wake.

\subsection{Carbon chemistry of the shock layer}
\label{sec:carbon}
\subsubsection{C$_2$ Swan and CH: early pyrolysis and boundary-layer chemistry}
This is the first identification of the C$_2$ Swan system in remote observations of a
sample-return capsule entry. Previous ground-based and airborne spectroscopy of the Stardust, Hayabusa and
Hayabusa2 capsules reported CN violet as the only carbon-bearing molecular emitter: the near-ultraviolet
spectrograph used for Stardust \citep{WinterTrumble2011} did not cover the Swan system, and the visible Stardust
spectra showed CN, the continuum and atomic lines \citep{JenniskensKoop2010,Jenniskens2010CN}; in the Hayabusa
spectra at these altitudes the capsule emission overlapped with that of the disintegrating spacecraft bus, which
prevented an analysis of the capsule's molecular bands there \citep{Abe2011}; and for Hayabusa2 the NASA
spectrographs covering 380--\SI{670}{nm} \citep{Prabhu2022} identified CN but not C$_2$. Hayabusa2, without a bus re-entry, therefore gives the first remote view of this phase
from a sample-return capsule. C$_2$ Swan has, however, been recorded in situ by an on-board spectrometer during the
orbital entry of the Varda W-2 capsule with a C-PICA heat shield, from about 98~km, together with \citet{Rao2026OSPREE}; there, too, C$_2$ and CH rose and peaked first and were attributed to the early pyrolysis of the
phenolic resin, while CN dominated deeper in the denser air. For Hayabusa, flow-field calculations with the
pyrolysis-gas chemistry (including C$_2$, CH and CN) predicted that molecular bands and atomic lines overlap the grey
body before the peak heating and fade quickly after it, but the bus fragments and the lack of sensitivity below
\SI{400}{nm} prevented a quantitative test \citep{Fujita2011AIAA}; the Hayabusa2 spectra, with C$_2$ above 69~km and CH above 72~km
and a declining CN-to-continuum ratio, show this trend. In natural meteors C$_2$ Swan has been reported only as
a probable identification in photographic fireball spectra \citep{Ceplecha1971,Borovicka1994lines}. In plasma
wind-tunnel tests of carbon-phenolic ablators, C$_2$ Swan is observed in the boundary layer in front of the
sample together with CN, CH, NH and H, in both air and nitrogen; it originates from the decomposition of the
pyrolysis hydrocarbons (C$_2$H$_2\rightarrow$C$_2$H$\rightarrow$C$_2$), is strongest at the beginning of a test
and decays as the outgassing declines, and it is absent for a non-pyrolysing carbon preform
\citep{Helber2013,Helber2014,Helber2016}. Laser ablation of a carbon--carbon composite produces C$_2$ in vacuum,
air and oxygen, whereas CN appears only when nitrogen is present \citep{Radhakrishnan2023}. Our detection of
Swan only above 69~km, at about 0.4 of the CN $\Delta v=-1$ band and with temperatures close to those of CN
($T_{\rm rot}\approx\SI{9200}{K}$, $T_{\rm vib}\approx\SI{7800}{K}$), fits this picture: C$_2$ marks the onset of
pyrolysis of the CFRP ablator at the beginning of the heating, when the boundary layer is rich in hydrocarbons,
and is excited in the same hot gas as CN. The CH A--X band, seen only at 80--72~km, belongs to
the same phase. The early CN maximum of Hayabusa near 80~km, associated with the onset
of pyrolysis \citep{Kraetzig2014}, and the high CN/continuum ratio at 80--72~km for Hayabusa2
(Section~\ref{sec:lines}) point to the same phase. The end of the Swan emission near 69~km, about
\SI{10}{s} after the first detection, does not, however, mark the end of pyrolysis: at 69~km the stagnation-point heat
flux is still rising towards its maximum (about \SI{7.6}{MW\,m^{-2}}, 60\,\% of the peak, in our model; peak at
58.5~km in the pre-flight prediction \citealp{Yamada2021}), so the resin continues to decompose in depth, and the
Balmer lines, which also trace the pyrolysis gas, keep rising until peak brightness (Section~\ref{sec:lines}). Two
effects should therefore be distinguished. (i) In-depth pyrolysis continues, but the gas reaching the surface changes:
early on it comes from virgin resin heated for the first time near the surface, whereas later the pyrolysis zone moves
inward and the gas percolates through the hot char layer, where hydrocarbons such as C$_2$H$_2$, the precursors of
C$_2$, are cracked to H$_2$ and carbon. (ii) The chemistry and excitation in the boundary layer change as the density
and the heat flux increase: more atomic N and O from the shock layer convert carbon to CN and CO, and collisional
quenching of the excited states grows. Our data cannot separate (i) and (ii); the persistence of CN and of the
hydrogen lines below 69~km shows that carbon and hydrogen are still injected, so the disappearance of C$_2$ reflects
a change in the composition of the injected gas and in the boundary-layer chemistry rather than the end of
pyrolysis. Below 69~km the carbon thus appears as CN rather than C$_2$, consistent with the ground tests in which CN
dominates the boundary-layer emission \citep{Helber2016}.

\subsubsection{CN}
The CN violet temperature of about \SI{7000}{K}, far above the surface temperature, shows that CN is
excited in the hot gas of the shock layer and boundary layer rather than near the wall temperature.
It is close to the \SI{8000 \pm 1000}{K} found for the Stardust capsule (PICA) at about 60~km
\citep{Jenniskens2010CN}, and lower than the \SI{13000}{K} that we derived with SPRADIAN from the mixed bus and capsule
spectra of Hayabusa \citep{Abe2011}, where N$_2^+$ of the shock layer of the fragmenting bus overlapped the CN bands. \citet{Kraetzig2014} found CN out of equilibrium
above 70~km and close to equilibrium after peak heating for Hayabusa; our free fits give
$T_{\rm rot}\approx T_{\rm vib}$ at 72--66~km, but they cannot test equilibrium below 66~km, so the single
temperature should be read as an effective one.

The CN band is strongest relative to the continuum at the highest altitudes and keeps a nearly constant
intensity from 78 to 62~km, while the continuum brightens by a factor of 8. For Hayabusa, a first CN maximum near 80~km was associated with the onset
of pyrolysis and a second, broader one with peak heating near 57~km \citep{Kraetzig2014}; for Hayabusa2 the
airborne CN signal also showed an early weak hump \citep{Prabhu2022}. This behaviour, and the decline of CN below
62~km while the continuum still increases, is consistent with CN produced from the
pyrolysis gas and by nitridation of the char, with oxidation to CO competing at lower altitudes
\citep{Helber2013,Helber2016}.

In contrast, CN is not detected in meteors even at high brightness: CN/Fe $<1/3$ and $<1/30$ in Leonids
\citep{Rairden2000,Jenniskens2004CN}, and no CN in the Bene\v{s}ov bolide \citep{BorovickaBerezhnoy2016}; in plasma wind
tunnel experiments CN appears only for carbon-rich meteorites and is strongest early in the ablation
\citep{Pisarcikova2023}. The strong CN emission of the capsule reflects the large carbon supply of the ablator and
the dense air of the low-altitude shock layer.

\subsubsection{N$_2^+$}
With the wavelength calibration and the second-order spectra, the near-ultraviolet peak lies at 388--\SI{389}{nm}, at the CN $\Delta v=0$ head, so CN is the main emitter, and
N$_2^+$(0,0) appears as the \SI{391}{nm} shoulder once the second-order resolution is fixed, at
$0.25\pm0.05$ of the CN $\Delta v=0$ band (66--55~km): N$_2^+$ first negative is present but about four times
weaker than CN. N$_2^+$
dominated the near-ultraviolet spectra of the Hayabusa entry above 83~km, with vibrational temperatures up to
\SI{13000}{K} \citep{Abe2011}, but that emission came mainly from the shock layer of the disintegrating
spacecraft bus; Hayabusa2 had no bus re-entry. In meteors, N$_2^+$ first negative was detected in a bright
Leonid fireball at 84~km and 72\,km\,s$^{-1}$ with a temperature of about \SI{10000}{K} and
N$_2^+$/N$_2\approx10^{-8}$ \citep{Abe2005}, and N$_2^+$ B--X was also identified in an ultraviolet meteor spectrum \citep{Abe2007ASR}, and meteor air-plasma temperatures depend on speed and mass
\citep{Jenniskens2004N2}; for Stardust (0.8~m, 12.9\,km\,s$^{-1}$), N$_2^+$ was observed with a rotational
temperature of about \SI{15000}{K} at 71.5--62~km \citep{WinterTrumble2011}. The weakness of N$_2^+$ relative to CN for Hayabusa2
below 80~km is consistent with (i) the lower speed and the small nose radius, which make the non-equilibrium
zone behind the shock, where N$_2^+$(B) is excited, very thin; (ii) collisional quenching of N$_2^+$(B): with a
quenching rate of about $9\times10^{-10}$\,cm$^3$\,s$^{-1}$ \citep{Dilecce2010} and a radiative rate of
$1.6\times10^7$\,s$^{-1}$, the critical density is $\sim2\times10^{16}$\,cm$^{-3}$, comparable to the post-shock
density below 70~km; and (iii) the dominance of CN, which is the main near-ultraviolet radiator of the Hayabusa
capsule at 51--53~km in radiation modelling \citep{Winter2019}. Low-resolution identifications of N$_2^+$(0,0) in
capsule spectra should consider the blend with the CN (0,0) head only \SI{3.1}{nm} away.

\subsection{Hydrogen, sodium and electron density}
\label{sec:hydrogen}
Balmer emission is clear in the capsule spectrum from 80 to 50~km at 11--8\,km\,s$^{-1}$. In meteors, hydrogen
lines belong to the hot ($\sim\SI{10000}{K}$) second spectral component, which appears only above about
30\,km\,s$^{-1}$ \citep{Borovicka1994,Matlovic2019,Matlovic2022,Vojacek2019,Vojacek2025}: H$\alpha$ was found in 62 of 304 video meteor spectra,
never below about 30\,km\,s$^{-1}$ \citep{Matlovic2022}, and the hot component of bright fireballs was modelled with
$T=\SI{10500}{K}$ and $N_e\approx5\times10^{19}$\,m$^{-3}$ \citep{Vojacek2025}. In meteors the hydrogen is attributed
to water and organics of the meteoroid, since atmospheric water at 80--90~km is far too scarce
\citep{JenniskensMandell2004}; OH A--X at \SI{309}{nm}, a direct tracer of water, has been searched for in ultraviolet
meteor spectra \citep{Abe2007ASR} (it lies below the range of our spectrograph). For the capsule, the hydrogen source is the pyrolysis gas of the phenolic resin of
the ablator, and the emission occurs in the dense shock and boundary layers at a much lower speed than in
meteors. The time history supports this: the hydrogen lines peak together with the continuum and fade faster
than it after peak heating, and the airborne observers found the H~I and Na~D signals peaking after the
shock-layer O~I and N~I lines, as expected for ablation products \citep{Prabhu2022}. Ablation of carbonaceous
meteorites in a plasma wind tunnel at about \SI{12}{km\,s^{-1}} equivalent conditions also produces H$\alpha$
\citep{Matlovic2022}. The H$\gamma$/H$\beta$ excitation temperature of 4300--\SI{6700}{K} is lower than the
shock-layer temperatures and the hot meteor component and close to the CN temperature, which suggests emission
from the boundary layer where the pyrolysis gas mixes with the shock-heated air (optical depth in H$\beta$
would bias the temperature upwards and cannot be excluded). The H$\beta$ Stark width gives a conservative upper limit of
$N_e<2\times10^{21}$\,m$^{-3}$ at 66--55~km for the hydrogen-emitting gas, which in Saha equilibrium corresponds
to temperatures below about \SI{8800}{K}; this, too, points to the boundary layer, and it provides a constraint for
flow-field simulations \citep{Prabhu2022}.

Na~D is strong throughout the observation (\SI{16}{W\,sr^{-1}} at peak brightness) and peaks with the continuum and
the Balmer lines. The ablator contains no sodium by design and the atmospheric sodium layer lies above 80~km, so the
sodium is most likely an impurity of the heat-shield material or a surface contamination released by ablation, in
line with the late peak of Na~D, after the shock-layer O~I and N~I lines, in the airborne spectra \citep{Prabhu2022}.
The probable Na~I 568.3/\SI{568.8}{nm} doublet at 52--49~km (Section~\ref{sec:lines}), which requires a strongly
self-absorbed Na~D, points to a large sodium column in a cool layer near the surface at the end of the heating.

\subsection{Surface temperature}
The colour temperature of 3000--\SI{3080}{K} reached below 64~km (\SI{3040 \pm 124}{K} at peak
brightness) lies within the range of the airborne Hayabusa2 results: \SI{2760 \pm 75}{K} at the peak of a
single-temperature fit \citep{Birch2022a}, about \SI{3200}{K} after correction for atmospheric
transmission \citep{Birch2022b}, and \SIrange{2500}{3700}{K} for the hottest part of a surface-temperature
distribution \citep{Ravichandran2025}. It is also similar to the Hayabusa capsule, which had the same type of
ablator and a similar entry speed: \SI{3100 \pm 300}{K} at about 50~km from the ground \citep{Ohnishi2011},
an apparent maximum of about \SI{3050}{K} from the aircraft \citep{Buttsworth2013}, and above \SI{3000}{K}
near peak heating at about 57~km \citep{Kraetzig2014}; at 42~km our ground grey-body temperature of the Hayabusa capsule was
\SI{2437}{K}, against a prediction of \SI{2525 \pm 50}{K} \citep{Abe2011}. The rise from \SI{2720}{K} at
70~km to \SI{3080}{K} at 64--60~km follows the increasing heat flux and the thermal response of the ablator.
The pre-flight prediction distributed for observation planning gives a stagnation-point wall temperature
of \SI{2364}{K} at 70.1~km, \SI{2673}{K} at 63.2~km and a maximum of \SI{2736}{K} at 60.3~km,
300--\SI{400}{K} below our colour temperatures but close to our brightness
temperatures (2260--\SI{2830}{K}) at the same altitudes; along the best-estimated
trajectory, a trajectory and material-response simulation for an assumed carbon-phenolic heat shield gives
a peak stagnation temperature of \SI{3119}{K} at 76~s after entry interface, i.e.\ at about 55~km, with
margins that make it an upper estimate \citep{Prabhu2022}. The peak heat flux occurs before the peak wall
temperature, the radiative component first \citep{Prabhu2022}. Our colour temperature reaches its maximum (\SI{3080}{K}) at
64--60~km, near the altitude of the predicted maximum wall temperature, and stays at about \SI{3010}{K} down to
52~km, past the altitude (about 55~km) at which the simulated stagnation temperature peaks; the brightness temperature, which also follows the
growing emitting area, keeps rising until 60--56~km. Our data begin earlier in the heating pulse than the airborne Hayabusa2
spectra, which cover the later part of the entry.

A single-temperature fit to a non-isothermal surface gives an effective temperature weighted towards the
hottest region and an effective area smaller than the visible area \citep{Birch2024}. Our emitting area of
0.2--0.6 of the projected forebody, increasing as the heating spreads over the forebody, therefore reflects both the emissivity of the charred CFRP and the
temperature distribution over the forebody, which cannot be separated with these data. The brightness
temperature, which uses the projected area and an emissivity of one, is a lower limit and lies
190--\SI{470}{K} below the colour temperature, consistent with an emissivity of order 0.4--0.6 for a
uniform surface or a hotter, smaller emitting region. Beyond 56~km the line of sight approaches the capsule from behind ($\theta>\SI{90}{\degree}$);
the continuum is then emitted by the conical flank of the forebody, which remains visible up to
$\theta=\SI{135}{\degree}$.

\subsection{Peak brightness and luminous efficiency}
\label{sec:peakdisc}
The radiant intensity in \SIrange{400}{680}{nm} peaks at 60.8~km, with an apparent V magnitude of $-5.0$
from Coober Pedy, similar to the maximum apparent magnitude of $-5.1$ from the campaign video systems
\citep{Sansom2022}. The altitude of the peak exposure rests on its own plate solution and the trajectory
geometry. It coincides with the altitude of the predicted maximum wall temperature (60.3~km) and is close to
the altitudes of the peak flux (about 56~km \citealp{Ohnishi2011}) and of the maximum luminous efficiency
(58.1~km \citealp{Sato2012}) of the Hayabusa capsule.

Our luminous efficiencies, $\tau_{400-680}\approx0.2$\,\% and $\tau_{\rm bol}=0.1$--3\,\% (Section~\ref{sec:lum}), refer
to a body of known mass, shape and speed on a known trajectory, and thus provide a calibration point for the
luminous efficiency of slow ($\sim$10\,km\,s$^{-1}$), large entering bodies, for which meteor data are scarce.
Two differences from the meteor definition should be kept in mind. In meteors the luminous efficiency refers to
the loss of kinetic energy including the ablated mass, $\frac12V^2{\rm d}m/{\rm d}t$, which dominates; for the
capsule this term is negligible (an ablated mass below about \SI{1}{kg} changes the energy loss by less than
10\,\%), so $\tau$ here is relative to the drag loss. And meteor values refer to a panchromatic or V passband:
$\tau_{400-680}$ is the quantity to compare with them, whereas $\tau_{\rm bol}$, dominated by the infrared
thermal emission of the surface, has no meteor counterpart. Unlike in
meteors, where much of the light comes from vaporised meteoroid material in the shock and the wake \citep{Abe2009LNP}, the capsule's
visible light is dominated by the thermal emission of the surface, and the persistent train adds only a faint
contribution (Section~\ref{sec:wake}); its luminous efficiency
is therefore set by the surface temperature and emitting area rather than by the ablated mass.
The Hayabusa capsule behaved in the same way. Its photographic light curve peaked at an absolute magnitude of $-3.5$
at 55--50~km, two magnitudes fainter than predicted, and with the meteor convention of \SI{1500}{W} radiated per
zero-magnitude its total radiated energy was \SI{2.8e5}{J}, about 0.03\,\% of the \SI{1e9}{J} deposited in the
atmosphere \citep{Borovicka2011}. The same convention gives \SI{5.2e5}{J} for our light curve (peak $-4.1$ at
60.8~km) over 80--49~km, 0.09\,\% of the kinetic energy lost over this interval (\SI{5.6e8}{J}) and 0.05\,\% of the kinetic energy at entry (the energy radiated at 400--\SI{680}{nm}, \SI{4.8e5}{J}, is 0.04\,\%; Section~\ref{sec:lum}). Measured, like the meteor values, against the
energy lost, both capsules are thus one to two orders of
magnitude less luminous than natural fireballs of similar energy, for which luminous efficiencies of 0.5--6\,\% are
expected, and than the disintegrating Hayabusa spacecraft (1.3\,\%) \citep{Borovicka2011}; even at peak brightness,
$\tau_{400-680}\approx0.2$\,\% stays below the fireball range: without a large envelope
of ablation vapour, the light comes from the hot surface only, and the low ablation coefficient of the capsule
($\sigma=0.0014$\,s$^2$\,km$^{-2}$ for Hayabusa \citealp{Borovicka2011}) reflects the same lack of mass loss.
A Long March 3B upper stage observed with the same method radiated 0.01\,\% of its kinetic energy at
420--\SI{645}{nm}, 99\,\% of it after its break-up into a cloud of burning fragments \citep{AbeLongMarch3}.

\subsection{Origin of the persistent train}
\label{sec:traindisc}
The train of the capsule (Section~\ref{sec:wake}) differs from the persistent trains of bright meteors, which last
minutes and are sustained by catalytic chemiluminescence of meteoric metals with ozone (Na, Fe with O$_3$) and spread
by diffusion and wind. Two properties constrain its origin. (i) The delayed onset, 0.5--\SI{0.9}{s} after the
passage, excludes thermal emission of the hot wake gas or of hot particles (soot, char), which would be brightest at the
capsule and fade monotonically; the emission must start as the wake gas cools and mixes with the ambient air.
(ii) The decay time does not follow the ambient density: between 66 and 50~km the number density $n$ of the air
rises by a factor of 6 (NRLMSISE-00), but $\tau$ stays at 0.9--3~s without a trend ($\tau\propto n^{0.03\pm0.10}$), whereas a
decay set by a two-body reaction or quenching with the ambient air ($\tau\propto1/n$) or by a three-body reaction such as
O + O$_2$ + M ($\tau\propto n^{-2}$) would shorten by factors of 6 and 36. The time scale is therefore probably set by
the cooling and the mixing of the wake gas with the ambient air, i.e.\ by the dynamics of the wake. A radiative lifetime would also give a constant $\tau$, but no radiative lifetime of the candidate
metastables fits, because O($^1$S) is quenched by O$_2$ within milliseconds at these densities. More generally,
forbidden lines cannot carry the train: the critical density of O($^1$S), where quenching by O$_2$ equals the radiative
rate, is about $5\times10^{13}$\,cm$^{-3}$ (about 90~km), 300--1000 times below the density at 72--53~km, so that at most
0.4--3\,\% of the O($^1$S) atoms emit at \SI{557.7}{nm} and the yields of O~I 630.0/\SI{636.4}{nm} and N~I
\SI{520}{nm} are $10^{-7}$ or less; this is why the green line of meteor afterglows and trains is seen only above
about 80~km \citep{Halliday1960,BorovickaJenniskens2000,Abe2004train}. The colour of the train, which could
discriminate between the remaining candidates, is not measured (Section~\ref{sec:wake}): chemiluminescence of the
cooling wake, such as the CO + O $\rightarrow$ CO$_2^*$ continuum (350--\SI{550}{nm}) expected in the wake of a carbon
ablator, where CO, the main gaseous product of char oxidation and pyrolysis, meets the atomic oxygen produced in the
shock layer, the NO + O $\rightarrow$ NO$_2^*$ air afterglow, or O$_2$ Herzberg emission from O-atom recombination,
remain possible. Of these, the NO + O $\rightarrow$ NO$_2^*$ afterglow is favoured by the delayed onset: its rate
coefficient is suppressed at high temperature by redissociation and quenching of the excited NO$_2$ and rises by a
factor of about 20 as the gas cools from 3000 to 1000~K \citep{ClyneThrush1962,Matsugi2025}, whereas the CO + O
$\rightarrow$ CO$_2^*$ emission has a positive activation energy (about 1900~K) \citep{ClyneThrush1962,SlackGrillo1985}
and weakens as the wake cools. Diffusion does not limit the visibility: molecular diffusion at 54--70~km
($D\approx0.03$--\SI{0.2}{m^2\,s^{-1}}) widens the train by less than \SI{1}{m} in \SI{2}{s}, far less than its width of about
\SI{100}{m}, which must therefore be set by the width of the wake when the train forms; the train fades because its chemistry ends, not because it disperses. Unlike meteoroids, the capsule deposits essentially no metal vapour, so the
long-lived metal--ozone cycle of meteor trains cannot operate and only this short afterglow remains. A spectrum
of the train would identify the emitter.

\subsection{Limitations}
\label{sec:limits}
\paragraph{Spatially unresolved emission} At the range of peak brightness (\SI{66}{km}) one pixel is about \SI{25}{m}
(21--\SI{68}{m} over the observed ranges), so every
spectrum integrates the shock layer, the boundary layer, the surface and the near wake (within about \SI{2}{km})
along the line of sight. The molecular and atomic temperatures are emission-weighted effective values, and their
assignment to regions of the flow (Figure~\ref{fig:schematic}) rests on their temperatures and time histories, not on
spatial information.

\paragraph{Resolution and band models} At \SI{2.7}{nm} the rotational structure is not resolved: $T_{\rm rot}$ is
constrained by the envelope of the band sequences and is correlated with $T_{\rm vib}$, the continuum and the
line-spread function, as the ranges of the systematic variants show (Section~\ref{sec:molecules}). The band models
assume optically thin emission and Boltzmann populations ($T_{\rm tr}=T_{\rm rot}$, $T_{\rm el}=T_{\rm vib}$).
Above about 70~km the shock layer of a \SI{0.4}{m} body at 11--\SI{12}{km\,s^{-1}} is expected to be in
thermochemical non-equilibrium, which the different rotational and vibrational temperatures of CN and C$_2$ at
80--72~km may reflect; below, the single temperature is an effective one. Self-absorption of the strong CN violet
bands in the boundary layer, if present, would flatten the band envelope and bias $T_{\rm rot}$ upwards; it cannot
be tested without resolved rotational lines.

\paragraph{Grey-body continuum} The colour temperature assumes an emissivity independent of wavelength over
475--\SI{608}{nm}. For an emissivity varying as $\lambda^{-n}$, the fitted temperature changes by about
$330\,n$~K at \SI{3000}{K} (Wien approximation over the span of the windows), i.e.\ by 30--\SI{100}{K} for
$n=0.1$--0.3; charred carbon composites are nearly grey in the visible, and this is within the systematic error.
Continuum emission of the gas (molecular band wings, free--bound and free--free emission) or of hot spalled
particles near the capsule would add to the surface emission; the agreement of the brightness temperatures in four
bands within \SI{90}{K} (Section~\ref{sec:continuum}) and the absence of a co-moving wake beyond about \SI{2}{km}
(Section~\ref{sec:wake}) argue against a large non-surface contribution, but cannot exclude it close to the
capsule.

\paragraph{Calibration} The absolute intensities depend on the transfer curve (0.5~mag between S-Log3 and Rec.~709) and on the
Sirius calibration (7\,\%); the colour temperature changes by 200--\SI{500}{K} between transfer curves (Table~\ref{tab:temp}).
The brightness temperature assumes that the whole visible forebody emits with the geometry of \citet{Prabhu2022}.

\paragraph{Trajectory and timing} The altitudes of our exposures follow from the observed directions and the trajectory geometry, and the
pre-flight trajectory agrees with the best-estimated trajectory \citep{Yamada2022BET} in altitude versus time
from entry interface (62.5 versus about 62.5~km at 70~s \citealp{Prabhu2022}). The UTC of our exposures,
however, is tied to the time epoch of the pre-flight trajectory (entry interface at 200~km at
17:27:39.5~UTC), because the camera clock was not synchronised. The campaign cameras give the first
detection at 103~km at 17:28:38.5~UTC \citep{Sansom2022}, 13~s later than the pre-flight trajectory
reaches that altitude, and the aircraft analyses report time offsets of several seconds between
measurement and the best-estimated trajectory \citep{Prabhu2022}. The altitudes and temperatures in this paper
do not depend on this epoch, but UTC values should be shifted if the entry-interface time of the
best-estimated trajectory differs from the pre-flight value; the time $t$ from the 80-km point, used in the figures, is not affected. %

\section{Conclusions}
\label{sec:conclusions}
We derived the continuum and molecular temperatures of the Hayabusa2 sample-return capsule from a
hand-tracked slitless video recorded with a consumer camera. A conical-diffraction grating model, a matched
extraction that removes motion smear, a transfer-curve test that identifies S-Log3, a calibration with Sirius
in the same exposures and a self-calibration of the second order give spectra with \SI{2.7}{nm} (first order)
and 2--\SI{3.5}{nm} (co-added second order) resolution from 80 to 49~km. The same reduction has also
been applied to the re-entry of a Long March 3B upper stage \citep{AbeLongMarch3}. The main results are:
\begin{enumerate}
\item Above 69~km the capsule spectrum is dominated by hot molecular emission of the shock layer. Besides the CN
violet system, the C$_2$ Swan bands are detected, for the first time in remote observations
of a sample-return capsule entry ($T_{\rm rot}\approx\SI{9200}{K}$, $T_{\rm vib}\approx\SI{7800}{K}$ at 80--72~km),
together with the CH A--X band. CN gives 6700--\SI{7200}{K} above 55~km, far above the surface temperature: it is
emitted by the hot gas of the shock and boundary layers. CH disappears below 72~km and C$_2$ near 69~km while CN
persists, which we relate to the early pyrolysis of the carbon-phenolic ablator, as in plasma wind-tunnel tests, and
to the change of the injected gas and of the boundary-layer chemistry while the heat flux is still rising (pyrolysis
continues below 69~km). N$_2^+$ first negative (0,0) is identified at \SI{391}{nm} but is four times weaker than CN
$\Delta v=0$.
\item Balmer lines, Na~D and O~I are detected. The H$\gamma$/H$\beta$ excitation temperature (4300--\SI{6700}{K})
and the H$\beta$ Stark limit on the electron density ($<2\times10^{21}$\,m$^{-3}$ at 66--55~km) show that hydrogen
from the pyrolysis gas radiates in the boundary layer.
\item The colour temperature of the heat shield rises from \SI{2720}{K} at 70~km to \SI{3080}{K} at 64--60~km and
stays at about \SI{3010}{K} below (\SI{3042 \pm 61 \pm 108}{K} at peak brightness), consistent with the airborne
results; the brightness temperature of the projected forebody rises from 2260 to \SI{2830}{K}.
\item The luminous efficiency relative to the loss of kinetic energy is about 0.2\,\% at 400--\SI{680}{nm} and
0.1--3\,\% (median 1.8\,\%) for the bolometric thermal emission; the peak brightness (60.8~km) precedes the
modelled peak heating (58.2~km) by \SI{2}{s}.
\item No co-moving wake is detected. The JAXA videos show a persistent train fixed in the air that lights up
0.5--\SI{0.9}{s} after the passage, peaks at 1.1--\SI{1.9}{s} and decays with $\tau=0.9$--\SI{3}{s} independently of
the ambient density. Its delayed onset and this independence exclude thermal emission and decay by reactions with the
ambient air, and point to chemiluminescence limited by the cooling and mixing of the wake gas.
\item For slitless spectra of carbon ablators, the CN B--X $\Delta v=-2$ sequence (\SIrange{439}{466}{nm}) must be
excluded from continuum fits, and the field-angle dependence of the infrared-cut filter edge must be corrected;
otherwise the continuum appears broader than a Planck curve and the colour temperature is biased by up to
\SI{1000}{K}.
\end{enumerate}

\section*{Declarations}
\section*{Availability of data and materials}
The extracted and calibrated spectra of every exposure, the fit results and the digest movie of
Appendix~\ref{app:movie} are openly available in the Zenodo repository \citep{AbeZenodo}. The JAXA videos of the fireball used in
Section~\ref{sec:wake} are available from the JAXA Hayabusa2 Project \citep{JAXAfireball2020} and are not redistributed.

\section*{Competing interests}
The authors declare that they have no competing interests.

\section*{Funding}
This work was supported by the 2026 Joint Research Program of the Aizu Research Center for Space Informatics (ARC-Space), University of Aizu.

\section*{Authors' contributions}
SA led the study, provided the observing equipment, analysed the data and wrote the manuscript. HARD and EKS prepared and coordinated the observations and carried out the field observations. M-YY prepared and coordinated the observations. KO contributed to the discussion of the CFRP ablator. ST, YI and KF prepared the observations, formed the JAXA field observation team and contributed to the discussion of the observational data. MY carried out the orbit determination of Hayabusa2 and coordinated the capsule observations. MCT and MC prepared the observations and carried out the field observations. TO and AU prepared the observations and analysed the data. SN, YH, TK, NS and YK prepared the observations. AC and GB coordinated the field observations. HD built the atmospheric entry database. All authors read and approved the final manuscript.

\acknowledgments{
We thank the Hayabusa2 project of JAXA and the members of the ground observation campaign. This work was supported by the 2026 Joint Research Program of the Aizu Research Center for Space Informatics (ARC-Space), University of Aizu.
}

\begin{figure}[tbp]
\centering
\includegraphics[width=\textwidth]{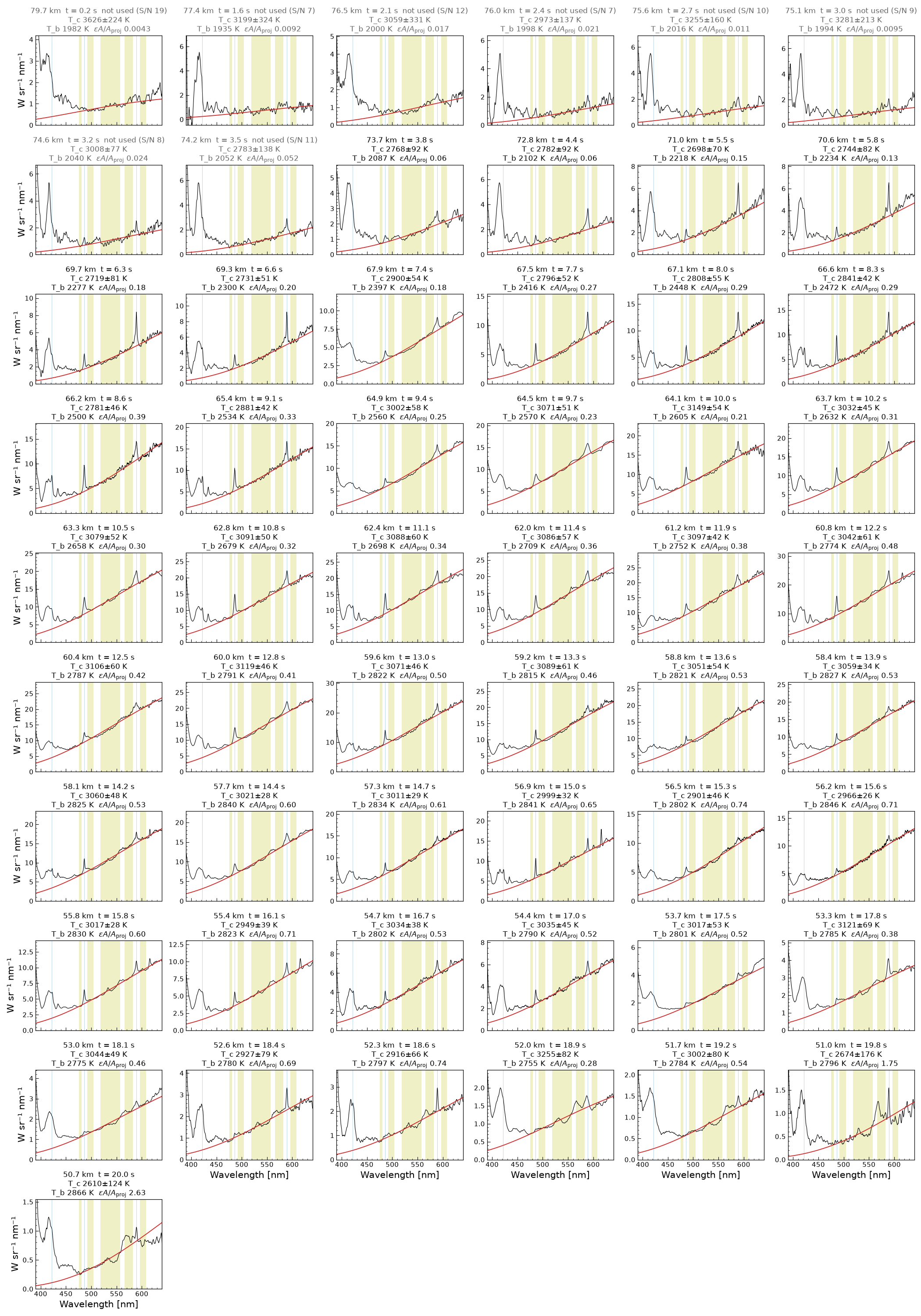}
\caption{\textbf{Grey-body fit to the spectrum of each exposure.} Spectrum of each exposure (black; S-Log3, Sirius calibration) with the grey-body fit (red); the first 8 panels (79.7--74.2~km, grey titles) are
fitted in the same way but not used for the temperature history (continuum S/N below 20). Yellow: fit windows; blue lines: CN \SI{388}{nm} and
\SI{422}{nm}, H$\beta$, Na~D. Panel titles: altitude, time $t$ in s (0~s at 80~km), colour temperature with
its statistical error, brightness temperature and $\epsilon A/A_{\rm proj}$.}
\label{fig:app_fits}
\end{figure}

\begin{figure}[tbp]
\centering
\includegraphics[width=\textwidth]{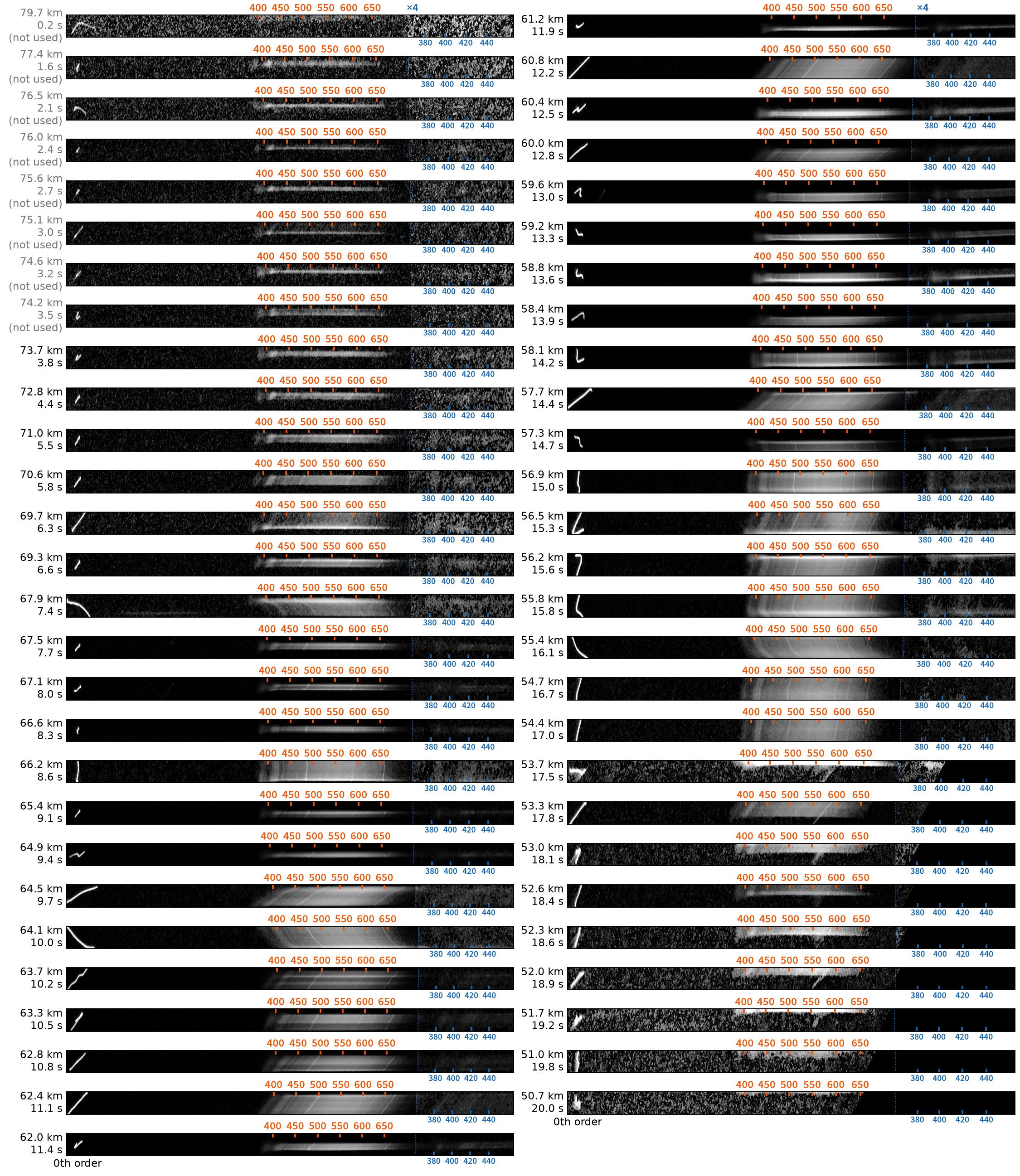}
\caption{\textbf{Image strips of all exposures along the dispersion direction.} Image strips of the exposures of Figure~\ref{fig:app_fits} (same order; altitude and time $t$ in s (0~s at 80~km) on the left;
grey labels: not used for the temperature history),
from the zero order (left) along the dispersion, $\pm18$~px across. Orange ticks: first-order wavelengths 400--\SI{650}{nm};
blue ticks: second-order wavelengths 380--\SI{440}{nm}; the second-order part (right of the dotted line, marked ``$\times4$'') is shown
four times brighter than the first order to make the weak second-order spectrum visible. Intensity: asinh stretch scaled to the first-order spectrum of each exposure; the zero
order is saturated.}
\label{fig:app_strips}
\end{figure}

\appendix{Image data and per-exposure fits}
\label{app:exposures}
Figure~\ref{fig:app_strips} shows the image data of all exposures used in this work: strips of the
background-subtracted, linearised exposures resampled along the dispersion direction from the capsule zero
order through the first-order spectrum to the second order. The shape of the zero-order image (the motion of
the capsule and of the hand-held camera during the 1/4-s exposure) sets the line-spread function of each
exposure through its autocorrelation along the dispersion (Section~\ref{sec:extraction}).
Figure~\ref{fig:app_fits} shows the calibrated first-order spectrum of each of the 47 exposures used for the
temperature history with its grey-body fit (Section~\ref{sec:continuum}), preceded by the 8 exposures between 79.7
and 74.2~km whose continuum signal-to-noise ratio (7--19) is below the selection threshold of 20; they are fitted
in the same way but not used for the temperature history, and they enter the 80--72~km bin of
Figure~\ref{fig:bgfits} and the molecular fits.

\appendix{Wavelength scale of each exposure}
\label{sec:wavecorr}
The matched extraction assigns wavelengths from the position of the zero-order image. The first-order
spectrum lies about 600 detector rows above the zero order (the second order up to about 1000 rows), and the
dispersed image does not always record the same part of the source's path during the exposure as the zero
order: in several exposures, mostly below 55~km, part of the zero-order streak has no dispersed counterpart, and
the image strips show a sharp edge parallel to the dispersion (Figure~\ref{fig:app_strips}). This is expected
from a readout-time offset between distant rows while the hand-tracked camera moves. The template then sits
off the dispersed image along the trace, which shifts the wavelength scale of that exposure. The shift is
measured in every exposure from the centres of H$\beta$, Na~D and H$\alpha$ (first-order counts; Gaussian of the
exposure's line-spread width on a quadratic continuum, centre free within $\pm\SI{5}{nm}$, lines above
5$\sigma$) and described as an offset along the trace, $\delta(s)=\alpha+\beta\,(s-s_{550})$, where $s$ is the
distance from the zero order and $s_{550}$ its value at \SI{550}{nm}. When the hydrogen lines are too weak to
constrain the slope (below 53.3~km), the strongest line gives $\alpha$ and $\beta$ is taken from the
neighbouring exposures. Below 53.3~km H$\alpha$ is not detected (below 3$\sigma$ also in co-adds of two to four
exposures), so the scale beyond \SI{600}{nm} cannot be checked there; at 420--\SI{590}{nm} it is confirmed by Na~D
($|\Delta\lambda|\le\SI{0.3}{nm}$) and by the free shifts of the CN and C$_2$ band fits (0.2--\SI{0.4}{nm} at 55--49~km). Before the correction, the line centres were within $\pm\SI{1}{nm}$ of the laboratory
values above 55~km, about \SI{-1}{nm} at 55--54~km and $-2$ to \SI{-4.5}{nm} (Na~D) below 53.3~km
($\alpha$ from $-6.4$ to \SI{-1.5}{px}). The template is shifted by $\delta(s)$ in the extraction of both
orders, and the measurement is iterated until the residual offsets converge (median $|\Delta\lambda|=\SI{0.05}{nm}$,
181 line measurements). The median slope ($\beta=0.0042$) is a scale error of the first-order grating model and is
not carried to the second order, whereas the per-exposure part is: with this model the second-order H$\beta$
centres measured before the correction are predicted to \SI{0.08}{nm} (median), against \SI{0.8}{nm} when the
full slope is carried. The correction changes the colour temperatures above 55~km by at most \SI{33}{K}
(median \SI{-4}{K}) and lowers those below 53.3~km by 50--\SI{170}{K}. The mean residual offsets are
$\delta\lambda=+0.09-0.0015\,(\lambda-500\,{\rm nm})$\,nm in the first order (H$\gamma$ $+0.08\pm0.07$, H$\beta$
$+0.15\pm0.04$, Na~D $-0.07\pm0.04$\,nm) and $-0.18-0.0117\,(\lambda-500\,{\rm nm})$\,nm in the second order
(H$\gamma$ $+0.59\pm0.12$, H$\beta$ $-0.02\pm0.08$\,nm); the first-order offset is removed in Figure~\ref{fig:altspec}, and the
band fits have a free wavelength shift. Extrapolated to \SI{390}{nm}, the second-order offset is $+1.1$\,nm,
as found in the N$_2^+$ fits (Section~\ref{sec:n2p}).

\appendix{Second-order resolution}
\label{sec:order2}
The second order has \SI{3.8}{px\,nm^{-1}}. For the band fits it is re-extracted with the same matched
kernel on a \SI{0.1}{nm} grid (\SIrange{375}{445}{nm}), and the measured line-spread function of every exposure
is kept. These line-spread functions differ strongly between exposures (FWHM \SIrange{0.5}{2.6}{nm}) because
of the motion smear along the dispersion, so co-adding exposures blurs the spectrum; the N$_2^+$ search below is
therefore made exposure by exposure. The second order is also broader than the measured line-spread function, by an amount that depends on
the brightness: the isolated H$\gamma$ line in 19 bright exposures (\SIrange{68}{52}{km}) requires an additional
Gaussian of \SI{2.0}{nm} FWHM (including the Stark width of H$\gamma$), whereas the faint spectra at 80--72 and
55--49~km show the vibrational structure of the CN and N$_2^+$ bands. We attribute the additional width to the
focus of the dispersed image and to the saturation of the zero-order core in the bright exposures, which makes the
zero-order kernel narrower than the true point-spread function. The resolution is therefore fitted for each
altitude bin: in the N$_2^+$ fits as an additional Gaussian common to the exposures of the bin (0, 1.5, 1.5, 2.5
and \SI{1.5}{nm} from 80--72 to 55--49~km), and in the CN fits of co-added spectra, which also contain the
spread of the line-spread functions and wavelength shifts between exposures, as the total FWHM (2.0, 3.0, 3.0,
3.5 and \SI{2.25}{nm}). Adjacent \SI{0.1}{nm} samples share pixels, and their noise is inflated by the square
root of the number of samples per pixel (2.6) in the fits.

\begin{figure}[tbp]
\centering
\includegraphics[width=\textwidth]{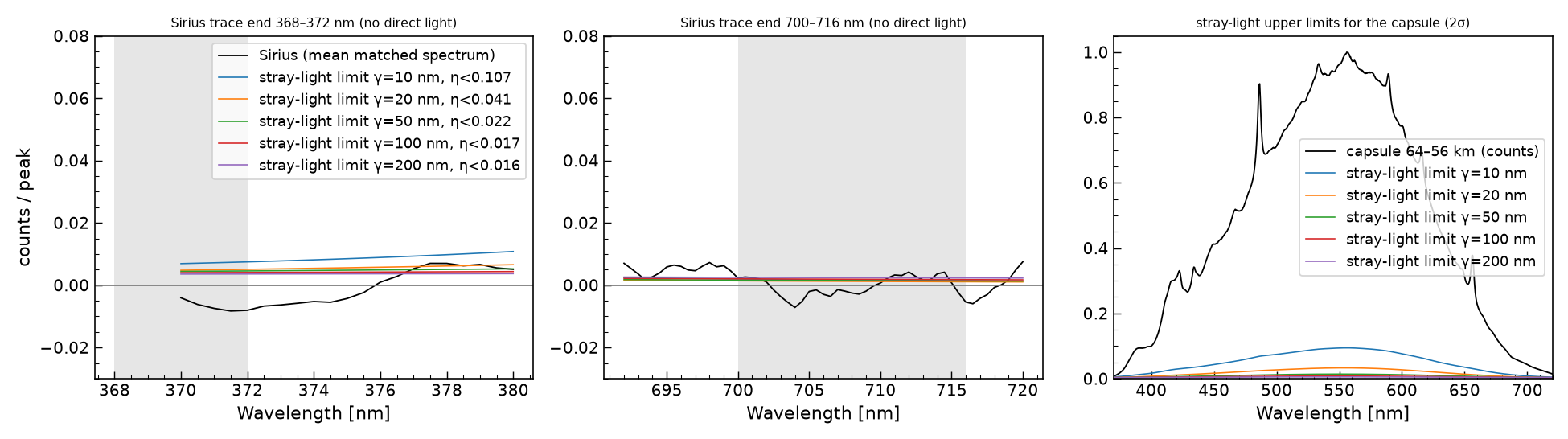}
\caption{\textbf{Stray-light model constrained with the Sirius spectra.} Stray-light model (Eq.~\ref{eq:stray}). Left and middle: mean matched Sirius spectrum at the ends of the
trace without direct light (grey bands), with the 2$\sigma$ upper limits of the scattered light for five kernel
widths. Right: the corresponding upper limits for the capsule spectrum at 64--56~km (counts, normalised).}
\label{fig:stray}
\end{figure}

\appendix{Stray light}
\label{sec:stray}
\label{app:stray}
Light of the dispersed spectrum may be scattered along the dispersion direction (grating scatter, lens halo)
and is then not removed by the background strips beside the trace. We model it as
\begin{equation}
C_{\rm obs}(\lambda) = C(\lambda) + \eta\,(K_\gamma * C)(\lambda),
\label{eq:stray}
\end{equation}
where $C$ are the counts of the direct light, $K_\gamma$ is a Lorentzian of half-width $\gamma$ (in wavelength along
the first order) normalised to unit area, and $\eta$ is the scattered fraction. The same instrument acts on
Sirius and on the capsule, so a correction $C=C_{\rm obs}-\eta\,K_\gamma*C_{\rm obs}$ would be applied to both
before the sensitivity is derived. The parameters are constrained by Sirius: at 368--\SI{372}{nm}, below the
cut-off of the lens, and at 700--\SI{716}{nm}, beyond the infrared-cut filter, its first order receives no direct
light, so the counts there are scattered light and noise. We use the matched (smear-free) Sirius spectra;
the aperture spectra are smeared by the trailing of Sirius and carry direct light into these ranges. For
every kernel width from 10 to \SI{200}{nm}, $\eta$ is consistent with zero ($0.03\pm0.04$, $0.00\pm0.02$,
$0.00\pm0.01$, $0.001\pm0.008$ and $0.003\pm0.007$ for $\gamma=10$, 20, 50, 100 and \SI{200}{nm}), and the fits do
not improve on $\eta=0$ (Figure~\ref{fig:stray}). Only kernels narrower than about \SI{7}{nm} are significant; they
describe the wings of the instrumental profile next to strong features, not scattered light, and they do not
change the shape of a smooth continuum. No stray-light correction is therefore applied. Applied to the capsule,
the 2$\sigma$ upper limits correspond to at most 10, 3.4, 1.5, 0.8 and 0.5\,\% of the signal at \SI{550}{nm}
for the five kernel widths. The calibration and the temperatures are repeated with each of these limits
applied to both Sirius and the capsule, and the largest change of the colour temperature is taken as the
stray-light systematic error (\SI{60}{K} at peak brightness, at most \SI{74}{K}).

\begin{figure}[tbp]
\centering
\includegraphics[width=\textwidth]{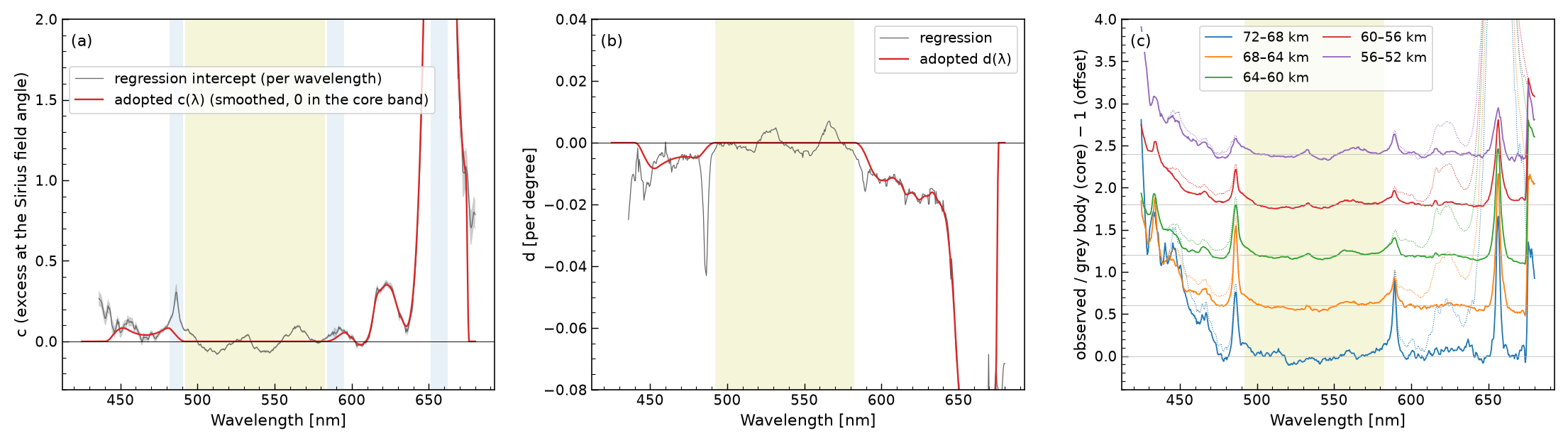}
\caption{\textbf{Signal-proportional sensitivity correction outside the core band.} Signal-proportional correction (Eq.~\ref{eq:selfcal}). (a) $c(\lambda)$: regression intercept at each
wavelength (grey, with its error) and the adopted smooth function (red), zero in the central band (yellow);
blue: masked lines. (b) $d(\lambda)$, the dependence on the field angle. (c) Median ratio of observed to
central grey-body intensity minus one in altitude bins, before (dotted) and after (solid) the correction
(offset for clarity).}
\label{fig:selfcal}
\end{figure}

\appendix{Signal-proportional correction outside the core band}
\label{app:selfcal}
\label{sec:selfcal}
Relative to a grey body fitted in the central part of the spectrum, the calibrated capsule spectra show an
excess below \SI{480}{nm} and above \SI{600}{nm} (Figure~\ref{fig:selfcal}). It is not an additive background:
beyond \SI{600}{nm} it is the same fraction of the signal (23--34\,\% at 615--\SI{625}{nm}) in altitude bins whose
brightness differs by a factor of 18, it rises abruptly at \SI{612}{nm}, on the edge of the infrared-cut filter,
and the counts at the ends of the trace are small (Appendix~\ref{sec:shape}). It decreases with the field angle
of the trace, by 1.2--2.2\,\% per degree at 600--\SI{640}{nm}, as expected for an interference edge that moves to
longer wavelengths towards normal incidence. Sirius, at 13--\SI{21}{\degree} from the optical axis and with few
counts beyond \SI{600}{nm}, determines the sensitivity there only to about $\pm30$\,\% per exposure. We
therefore describe the excess by a factor
\begin{equation}
F_e(\lambda)=1+w(\lambda)\left[c(\lambda)+d(\lambda)\,\{\theta_e(\lambda)-\theta_S(\lambda)\}\right],
\label{eq:selfcal}
\end{equation}
where $\theta_e$ is the field angle of the trace of exposure $e$ and $\theta_S$ the mean for the Sirius
exposures. The functions $c$ and $d$ are measured from the 48 exposures with a continuum signal-to-noise ratio of
at least 20 in the uncorrected spectra (47 after the correction, Section~\ref{sec:continuum}). Each is fitted with a grey body in the central windows (492--504.5, 518.5--557.5 and
566--\SI{582.5}{nm}), and at every wavelength the ratio of observed to grey-body intensity minus one is regressed on
$(1,\ \theta_e-\theta_S,\ P_e)$, where $P_e$, the CN $\Delta v=-1$ band relative to the continuum at \SI{550}{nm},
takes up molecular emission and is not part of the correction. H$\beta$, Na~D and H$\alpha$ are masked and $c$
and $d$ interpolated across them. Above the central band $c$ and $d$ are smoothed with a \SI{1.5}{nm} Gaussian;
below it they are quadratic in wavelength (fitted at 445--\SI{482}{nm}), so that the band structure of CN
$\Delta v=-2$ and Swan $\Delta v=+1$ stays in the spectrum. The weight $w$ is zero in the central band, which sets
the temperature scale, and rises smoothly to one over \SI{14}{nm} on both sides; it falls to zero between 455 and
\SI{440}{nm}, so that H$\gamma$ and the bands below \SI{440}{nm} are not changed, and above \SI{676}{nm}. At the
Sirius field angle, $c$ is 0.04--0.11 at 450--\SI{480}{nm}, 0.01--0.02 at 600--\SI{610}{nm}, 0.2--0.34 at
615--\SI{640}{nm} and about 4 at H$\alpha$; $d$ is $-0.005$ to $-0.010$ per degree below \SI{480}{nm} and
$-0.012$ to $-0.022$ per degree at 600--\SI{640}{nm}. The capsule spectra are divided by $F_e$. Afterwards the
residuals with respect to the central grey body are within $\pm5$\,\% at 480--\SI{640}{nm} in the altitude bins
below 68~km (8\,\% at 72--68~km); the instrumental part of the structure at 612--\SI{616}{nm}, which is
also seen in the Sirius ratio (about $+6$\,\% at \SI{616}{nm}), is removed, and the radiant intensity at
640--\SI{680}{nm}, previously inflated by the underestimated sensitivity, decreases by a factor of 2--5, which
halves the peak radiant intensity integrated over 400--\SI{680}{nm}. At \SI{616}{nm} the intercept $c$ exceeds its
neighbours by about 11\,\% of the continuum, about twice the Sirius structure, so that the correction also takes up
part of the O~I \SI{615.7}{nm} emission (Section~\ref{sec:lines}). Below \SI{470}{nm} the residuals that remain
at high altitude are molecular emission. Without the correction the colour temperature at peak brightness is
\SI{2968}{K} (\SI{-74}{K}).

\subsection*{The shape of the continuum}
\label{sec:shape}
In earlier versions of this analysis the continuum was fitted between 440 and \SI{608}{nm}. The calibrated
spectra then appear broader than a Planck curve: the colour temperature depends on the windows by several
hundred kelvins, and an additional pedestal, constant in detector counts and free in amplitude, of 17--43\,\%
of the signal at \SI{550}{nm} was needed, which gave about \SI{2850}{K} at peak brightness. Three checks show
that this pedestal was not a property of the instrument or of the continuum. (i) The stray-light model of
Appendix~\ref{app:stray}, constrained by the Sirius spectrum where no direct light falls, allows at most a few
per cent of scattered light (\SI{0.5}{\%} to \SI{10}{\%} at \SI{550}{nm} for kernel widths of 200 to \SI{10}{nm}). (ii) In
detector counts, the fitted pedestal (about 2900 counts at 64--56~km) exceeds the total signal at the ends of
the capsule trace (about 1000 counts at 392--402 and 650--\SI{690}{nm}, 140 at \SI{370}{nm} and 250 at
\SI{720}{nm}), and it absorbs weak emission features inside the windows. (iii) The excess is concentrated at
440--\SI{466}{nm}, where the CN B--X $\Delta v=-2$ sequence lies; a free CN component there would need 5--10
times the $\Delta v=-2/\Delta v=-1$ ratio predicted by METEORAD (about 0.1), and C$_2$ Swan does not account for
it, so the CN band and possibly other weak emission make this range unsuitable for the continuum. (iv) What remains outside the central band after excluding this range is proportional to the signal and
depends on the field angle (Appendix~\ref{sec:selfcal}): it is a property of the sensitivity at the edges of the
pass band, which Sirius, at larger field angles and with few counts there, does not determine, and not an
additive background. Without the CN range and with this correction,
a grey body alone fits the windows (reduced $\chi^2\approx0.6$), the colour temperature is nearly independent
of any pedestal (\SI{3042}{K} without and \SI{3043}{K} with a pedestal bounded by the end counts at peak
brightness). A laboratory measurement of the instrumental
scattered light would further tighten the limits of Appendix~\ref{app:stray}.

\appendix{Temperatures of the emission models}
\label{app:temps}
The model spectra describe each degree of freedom of the emitting gas by its own temperature, as in SPRADIAN
\citep{FujitaAbe1997,FujitaAbe2003}: translational ($T_{\rm tr}$), rotational ($T_{\rm rot}$), vibrational
($T_{\rm vib}$) and electronic excitation ($T_{\rm el}$); the surface is described by a colour and a brightness
temperature. All emission is taken as optically thin.

\paragraph{Translational temperature}
$T_{\rm tr}$ sets the Doppler width of a line of a species of mass $M$,
\begin{equation}
\Delta\lambda_{\rm D}=\frac{\lambda}{c}\sqrt{\frac{8\ln2\,kT_{\rm tr}}{M}},
\end{equation}
about \SI{0.005}{nm} for CN at \SI{7000}{K}, far below the line-spread function; it is set to $T_{\rm rot}$.

\paragraph{Electronic excitation temperature}
The population of an atomic level $u$, or of an electronic state of a molecule, of energy $E_u$ and statistical
weight $g_u$ is
\begin{equation}
n_u=N\,\frac{g_u\exp(-E_u/kT_{\rm el})}{U(T_{\rm el})},\qquad U(T)=\sum_j g_j\exp(-E_j/kT),
\label{eq:boltz}
\end{equation}
and a line emits $I_{ul}=(hc/4\pi\lambda_{ul})A_{ul}n_uL$ per unit area and solid angle along the path $L$. The
excitation temperature of Eq.~\ref{eq:texc} is $T_{\rm el}$ of hydrogen derived from two lines with this
relation; the ionization balance at the same temperature is given by the Saha equation (Eq.~\ref{eq:saha}).

\paragraph{Vibrational temperature}
Within an electronic state, the vibrational levels $v$ are populated as
\begin{equation}
\begin{split}
n_v&=n_{\rm el}\,\frac{\exp\!\left[-hc\,G(v)/kT_{\rm vib}\right]}{Q_{\rm vib}},\\
G(v)&=\omega_e\!\left(v+\tfrac12\right)-\omega_ex_e\!\left(v+\tfrac12\right)^2,
\end{split}
\end{equation}
with the term values $G(v)$ in wavenumber units and $Q_{\rm vib}$ the sum of the numerators.

\paragraph{Rotational temperature}
Within a vibrational level, the rotational levels $J$ are populated as
\begin{equation}
\begin{split}
n_{vJ}&=n_v\,\frac{(2J+1)\exp\!\left[-hc\,F_v(J)/kT_{\rm rot}\right]}{Q_{\rm rot}},\\
F_v(J)&=B_vJ(J+1)-D_vJ^2(J+1)^2,
\end{split}
\end{equation}
and a rotational line $v'J'\rightarrow v''J''$ emits
\begin{equation}
I=\frac{hc}{4\pi\lambda}\,A_{v'v''}\,\frac{S_{J'J''}}{2J'+1}\,n_{v'J'}\,L,
\end{equation}
with the band transition probability $A_{v'v''}$ and the H\"onl--London factor $S_{J'J''}$. The molecular fits
(Sections~\ref{sec:cn} and \ref{sec:swan}) vary $T_{\rm rot}$ and $T_{\rm vib}$ independently, with
$T_{\rm tr}=T_{\rm rot}$ and $T_{\rm el}=T_{\rm vib}$; the relative band shapes at a resolution of a few nanometres are
set mainly by $T_{\rm vib}$ (the band sequence) and by $T_{\rm rot}$ (the extent of each band towards its tail).

\paragraph{Colour and brightness temperatures of the surface}
The continuum is the grey body of Eq.~\ref{eq:model} with the Planck function
\begin{equation}
B(\lambda,T)=\frac{2hc^2}{\lambda^5}\,\frac{1}{\exp(hc/\lambda kT)-1};
\end{equation}
its shape gives the colour temperature $T_c$, and the intensity in a band relative to the projected forebody
gives the brightness temperature $T_b$ (Eq.~\ref{eq:tb}), which equals the surface temperature only for an
emissivity of one and a uniform surface.

\appendix{N$_2^+$ first negative system: fits and alternatives}
\label{app:n2p}
Each exposure is fitted with METEORAD spectra of CN $\Delta v=0$ (at the CN temperature of
Section~\ref{sec:cn}) and of N$_2^+$ (0,0)+(1,1), both convolved with that exposure's line-spread function and
the focus width of its altitude bin (Appendix~\ref{sec:order2}), plus a linear continuum and a wavelength shift;
the band ratio N$_2^+$(0,0)/CN($\Delta v=0$) of the exposures in a bin is averaged with inverse-variance weights,
the error being inflated by the scatter. With the resolution fixed in this way, CN alone does not reproduce the
red side of the \SIrange{388}{392}{nm} profile, and adding N$_2^+$ lowers the summed $\chi^2$ from 1220 to 598
(66--60~km, 1572 samples) and from 1108 to 590 (60--55~km, 1703 samples). The ratio is $0.24\pm0.07$ at
66--60~km and $0.26\pm0.06$ at 60--55~km, and $0.14\pm0.12$ and $0.20\pm0.13$ at 72--66 and 55--49~km
($T_{\rm N_2^+}=\SI{7000}{K}$; 2$\sigma$ upper limits 0.37--0.47; Figure~\ref{fig:n2p}); at 80--72~km, where the
vibrational structure of CN $\Delta v=0$ is resolved without additional focus width, N$_2^+$ is not detected
($<0.42$). Below 72~km the ratio changes by less than 0.02 when the focus width is varied by $\pm\SI{0.5}{nm}$
(at 80--72~km the limit stays at 0.42--0.43), and it increases
by about 25\,\% for $T_{\rm N_2^+}=\SI{10000}{K}$. Taking the wavelength shift from the sharp CN $\Delta v=-1$
heads at 418--\SI{422}{nm} instead of fitting it raises the ratio to about 0.5 but fits much worse ($\chi^2$ 2014
versus 598 at 66--60~km), which indicates an offset of the second-order wavelength scale near \SI{390}{nm}
relative to \SI{420}{nm} (Appendix~\ref{sec:wavecorr}). Two atomic alternatives were tested at 66--60~km, where
the bands are best resolved among the bright bins: first-order O~I \SI{777}{nm}, which falls on second-order
\SI{388.7}{nm}, gives $\chi^2=722$ and an atomic line at \SI{390.55}{nm} (Si~I) gives 816 (for 14 instead of 12 fitted exposures;
0.44 versus 0.38 per sample), both worse than N$_2^+$ (598); at 60--55~km, with a broader focus, the three are comparable (568--609). The first-order
sensitivity at \SI{777}{nm} is strongly suppressed by the infrared-cut filter and the CFRP ablator offers no
silicon source, so we attribute the \SI{391}{nm} excess to N$_2^+$ B--X (0,0)+(1,1). Combining 66--60 and
60--55~km, where the second order is brightest, gives N$_2^+$(0,0)/CN($\Delta v=0$) $=0.25\pm0.05$
($0.31\pm0.06$ for $T_{\rm N_2^+}=\SI{10000}{K}$). The absolute ratio relies on the extrapolation of the
second-order sensitivity below \SI{395}{nm} (the two bands are \SI{3}{nm} apart) and on the second-order
resolution.

\appendix{Weak and unidentified features}
\label{app:features}
The remaining features (Figure~\ref{fig:altspec}) were examined with METEORAD spectra of 22 air and 20 metal and
oxide species at 3000--\SI{12000}{K}, convolved with the measured line-spread function of each altitude bin (including
its far wings): each species was added in turn to a model of the identified emitters, and the position of an
additional line was scanned in windows around each feature. Line intensities are measured as for H$\beta$ (a Gaussian
of the line-spread width on a linear continuum; with the measured line-spread function the O~I \SI{615.7}{nm}
multiplet gives 4.0 and \SI{5.2}{W\,sr^{-1}} at 66--60 and 60--55~km). (i) At 66--55~km O~I \SI{615.7}{nm} (3p\,$^5$P--4d\,$^5$D$^\circ$)
remains after the correction of Appendix~\ref{sec:selfcal} with 1.9 and \SI{2.4}{W\,sr^{-1}} at 66--60 and 60--55~km
(free line at 615.85--\SI{616.0}{nm}), together with O~I \SI{533.0}{nm} (2.1 and \SI{1.4}{W\,sr^{-1}}, at
\SI{533.2}{nm}); at 60--55~km O~I lowers $\chi^2$ more than any other candidate. The ratio 533.0/615.7 (1.1 and 0.6)
exceeds the optically thin LTE value (0.45--0.49 for 4500--\SI{12000}{K}), but the correction also takes up part of
the \SI{615.7}{nm} line, which cannot be separated from the excess at the filter edge; the \SI{615.7}{nm} fluxes are
therefore lower limits. Na~I 615.4/\SI{616.1}{nm} is too weak: below $10^{-3}$ of Na~D in optically thin LTE, and about 0.16 of the
Na~I \SI{568}{nm} doublet, which is not detected above 52~km (see (iii)). At 72--66~km and below 55~km the
\SI{533}{nm} feature is stronger than O~I alone would give (ratios 3.3 and 1.5), and below 55~km it lies at
\SI{532.3}{nm}, so part of it remains unexplained. (ii) The feature at
448--\SI{449}{nm} (80--72 and 55--49~km; free line at \SI{449.8 \pm 0.2}{nm} at 80--72~km) is part of the excess in the CN $\Delta v=-2$ range
(Appendix~\ref{sec:shape}); METEORAD places bands of high vibrational levels of this sequence at \SI{447}{nm}, but
weaker than observed. Mg~II \SI{448.1}{nm} lies \SI{1.7}{nm} below the observed position and needs temperatures above
\SI{10000}{K}, and the ablator contains no magnesium. (iii) At 52--49~km a line at \SI{568.15}{nm} (568.05--\SI{568.3}{nm}) is detected
(\SI{0.36 \pm 0.03}{W\,sr^{-1}}, 0.21 of Na~D, $\Delta\chi^2=181$; Gaussian with the width of Na~D on a quadratic baseline;
a line with the measured line-spread function is found at \SI{568.30 \pm 0.07}{nm}); in the
higher bins no line is found at 567--\SI{569}{nm}. Its position matches the Na~I 568.3/\SI{568.8}{nm} subordinate doublet
(4d--3p), and the other subordinate lines are consistent with it (Na~I 615.4/\SI{616.1}{nm}, expected at about
0.16 of the doublet in LTE, i.e.\ \SI{0.06}{W\,sr^{-1}}: measured \SI{-0.02 \pm 0.06}{W\,sr^{-1}}; \SI{498.3}{nm} not detected). In optically thin LTE,
however, the doublet would be only $7\times10^{-5}$ to $5\times10^{-3}$ of Na~D at 3000--\SI{6000}{K}, so that Na~D
would have to be self-absorbed by a factor of 40--3000, e.g.\ in a cool ($\sim$\SI{3000}{K}), sodium-rich layer
near the surface; the identification therefore remains probable rather than certain, and non-LTE excitation
cannot be excluded. N~II \SI{567.9}{nm} would need temperatures far above those reached at 8--\SI{9}{km\,s^{-1}},
Cu~I \SI{570.0}{nm} is excluded by the absence of the stronger Cu~I 510.6, 515.3, 521.8 and \SI{578.2}{nm} lines,
Fe~I by the absence of 527.0 and \SI{537.1}{nm}, and C$_2$ by the absence of the Swan bands below 69~km.
(iv) Ca~I \SI{422.7}{nm}, included in the CN fits (Section~\ref{sec:molecules}), is seen as a separate line only
below 55~km: at 55--52~km a free line lies at \SI{422.70 \pm 0.05}{nm} (\SI{1.4}{W\,sr^{-1}}, $\Delta\chi^2=198$),
whereas at 72--55~km the free line falls at 422.15--\SI{422.6}{nm}, on the CN head at \SI{421.6}{nm}, from which it
cannot be separated. (v) The residual at 585--\SI{587}{nm} is the wing of Na~D, which is broader than the measured
line-spread function in every bin (an additional Gaussian of 1.25--\SI{3.5}{nm} FWHM, consistent with the instrumental
width $\sigma_x$ fitted on Na~D for the Stark width below); the free line moves from 585.4 to \SI{587.4}{nm} between the bins, so it is
not a separate line. (vi) At 55--49~km a weak line at \SI{636.7 \pm 0.2}{nm} (\SI{0.30}{W\,sr^{-1}}, $\Delta\chi^2=64$)
could be Si~II \SI{637.1}{nm} or the N$_2$ first positive system; it lies where the correction of
Appendix~\ref{sec:selfcal} is large and is not seen consistently in the other bins, so it is not identified.

\appendix{Stagnation-point heating model}
\label{app:heating}
The heating rate of Figure~\ref{fig:heating}c is a cold-wall estimate at the stagnation point from engineering
correlations; it is not a flow-field or material-response computation.

\paragraph{Trajectory and atmosphere} The positions of the pre-flight trajectory (Earth-fixed) are interpolated with a
cubic spline, and the speed $V$ is the magnitude of their time derivative, i.e.\ the speed relative to the rotating
atmosphere. The model is evaluated every \SI{0.1}{s} between 100 and 30~km. The air density $\rho$ is taken from
NRLMSISE-00 \citep{Picone2002} at the time, latitude, longitude and altitude of each point, with
$F_{10.7}=85$ (daily and 81-day mean) and $A_p=4$. The nose radius is $R_n=\SI{0.207}{m}$ \citep{Prabhu2022}.

\paragraph{Convective heating} The adopted convective rate is the Detra--Kemp--Riddell correlation
\citep{DetraKempRiddell1957},
\begin{equation}
q_{\rm c}=\frac{C_{\rm DKR}}{\sqrt{R_n}}\left(\frac{\rho}{\rho_{\rm SL}}\right)^{1/2}\left(\frac{V}{V_{\rm c}}\right)^{3.15},
\label{eq:dkr}
\end{equation}
with $C_{\rm DKR}=1.1035\times10^{8}$~W\,m$^{-3/2}$ (17600~Btu\,ft$^{-3/2}$\,s$^{-1}$), $\rho_{\rm SL}=\SI{1.225}{kg\,m^{-3}}$
and $V_{\rm c}=\SI{7.95}{km\,s^{-1}}$. For comparison, the Sutton--Graves relation \citep{SuttonGraves1971},
\begin{equation}
q_{\rm c,SG}=k\sqrt{\rho/R_n}\,V^3,\qquad k=1.7415\times10^{-4}~{\rm kg^{1/2}\,m^{-1}},
\end{equation}
gives a peak of \SI{9.0}{MW\,m^{-2}} at 56.3~km; at the peak of $q_{\rm c,SG}$, Eq.~\ref{eq:dkr} is 18\,\% higher.

\paragraph{Radiative heating} The equilibrium radiative rate is the Tauber--Sutton correlation for air
\citep{TauberSutton1991},
\begin{equation}
q_{\rm r}=C\,R_n^{\,a}\,\rho^{1.22}\,f(V)\quad[{\rm W\,cm^{-2}}],\qquad
a=\min\!\left(1.072\times10^{6}\,V^{-1.88}\rho^{-0.325},\,1\right),
\label{eq:ts}
\end{equation}
with $C=4.736\times10^4$, $R_n$ in m, $\rho$ in \si{kg\,m^{-3}} and $V$ in \si{m\,s^{-1}}. The velocity function
$f(V)$ is interpolated linearly in Table~\ref{tab:fV} and set to zero below \SI{9}{km\,s^{-1}}. The correlation
is stated for 10--\SI{16}{km\,s^{-1}} and $R_n=0.3$--\SI{3}{m}; below \SI{10}{km\,s^{-1}} (under 56~km) it is
extrapolated (dotted in Figure~\ref{fig:heating}c), and the nose radius of the capsule is below its range.

\begin{table}[!htbp]
\centering\small\setlength{\tabcolsep}{3.5pt}
\caption{\textbf{Velocity function $f(V)$ of the Tauber--Sutton radiative heating correlation.} (Eq.~\ref{eq:ts}) used in this work.}
\label{tab:fV}
\begin{tabular}{cccccccccc}
\toprule
$V$ (\si{km\,s^{-1}}) & 9.0 & 9.25 & 9.5 & 9.75 & 10.0 & 10.25 & 10.5 & 10.75 & 11.0\\
$f(V)$ & 1.5 & 4.3 & 9.7 & 19.5 & 35 & 55 & 81 & 115 & 151\\
\midrule
$V$ (\si{km\,s^{-1}}) & 11.5 & 12.0 & 12.5 & 13.0 & 13.5 & 14.0 & 14.5 & 15.0 & 15.5\\
$f(V)$ & 238 & 359 & 495 & 660 & 850 & 1065 & 1313 & 1550 & 1780\\
\midrule
$V$ (\si{km\,s^{-1}}) & 16.0 &&&&&&&&\\
$f(V)$ & 2040 &&&&&&&&\\
\bottomrule
\end{tabular}
\end{table}

\paragraph{Total and limitations} The plotted total is $q=q_{\rm c}+q_{\rm r}$. Table~\ref{tab:heat} lists the model
along the trajectory. Not included are the reduction of the convective heating by the wall enthalpy (a few per cent
for a \SI{3000}{K} wall at a total enthalpy of about \SI{60}{MJ\,kg^{-1}}), by the blowing of pyrolysis and ablation
gas, and by radiative cooling of the shock layer, as well as non-equilibrium radiation and the angle of attack. The
model thus serves to locate the heating pulse in time and altitude; its magnitude agrees with the pre-flight
prediction \citep{Yamada2021} (Section~\ref{sec:lum}).

\begin{table}[!htbp]
\centering\small\setlength{\tabcolsep}{3.5pt}
\caption{\textbf{Stagnation-point heating model along the pre-flight trajectory.} (Eqs.~\ref{eq:dkr} and \ref{eq:ts};
$R_n=\SI{0.207}{m}$, NRLMSISE-00). $t$: time after the capsule passed 80~km (17:28:38.4~UTC); $a$: exponent in Eq.~\ref{eq:ts}; heating rates $q_{\rm c}$, $q_{\rm r}$ and $q=q_{\rm c}+q_{\rm r}$ in \si{MW\,m^{-2}}.}
\label{tab:heat}
\begin{tabular}{rrcccccc}
\toprule
$h$ (km) & $t$ (s) & $V$ (\si{km\,s^{-1}}) & $\rho$ (\si{kg\,m^{-3}}) & $a$ & $q_{\rm c}$ & $q_{\rm r}$ & $q$\\
\midrule
80.0 & 0.0 & 11.63 & \num{2.1e-5} & 0.81 & 3.3 & 0.07 & 3.4\\
72.0 & 4.9 & 11.50 & \num{6.8e-5} & 0.56 & 5.8 & 0.38 & 6.1\\
69.0 & 6.8 & 11.41 & \num{1.0e-4} & 0.50 & 6.9 & 0.65 & 7.6\\
66.0 & 8.7 & 11.26 & \num{1.5e-4} & 0.45 & 8.1 & 0.99 & 9.0\\
60.8 & 12.2 & 10.82 & \num{3.0e-4} & 0.39 & 10.0 & 1.58 & 11.5\\
58.2 & 14.2 & 10.44 & \num{4.1e-4} & 0.38 & 10.5 & 1.46 & 12.0\\
55.0 & 16.7 & 9.81 & \num{6.1e-4} & 0.37 & 10.5 & 0.73 & 11.2\\
49.0 & 22.0 & 7.87 & \num{1.3e-3} & 0.44 & 7.6 & 0.00 & 7.6\\
\bottomrule
\end{tabular}
\end{table}

\appendix{Timing and calibration of the JAXA videos}
\label{app:jaxa}
The time of the fixed-camera JAXA video (Section~\ref{sec:wake}) was fixed by matching the light curve of the capsule (head flux within 12~px, below the
saturated maximum) to our spectroscopic light curve $I_{400-680}$: the best offset is determined to
$\pm$\SI{0.2}{s}. The video was calibrated on the stars. Its field (about \SI{12.8}{px\,deg^{-1}} at the centre,
\ang{150} across) was solved with a fish-eye projection on 115 catalogue stars (median residual 0.4~px; seeded with
Sirius and the stars of Orion), and aperture photometry of the 36 unsaturated stars with peaks above 10~DN
($V=0.4$--5.3) on the mean of the frames before the passage fixes the tone curve: with the pixel values linearised
as $(\mathrm{DN}/255)^{\gamma}$, the stellar fluxes follow $V$ with a slope of 1 for $\gamma=0.9$ (rms 0.11~dex; we adopt
$\gamma=0.9$ and give $\gamma=1$ as the alternative). The video is thus close to linear in this range; the slope of the
saturated head flux against $I$ ($\log F_{\rm video}\propto0.38\log I$) reflects the saturation, not the encoding,
and a gamma derived from it ($\approx2.7$) would only be an upper bound.
Fainter point sources are partly erased by the compression of the published video. The directions of the capsule
through the plate solution, compared with the predicted trajectory, place the camera near Coober Pedy
(29.005$^\circ$\,S, 134.693$^\circ$\,E), at 67--\SI{77}{km} from the capsule at 66--54~km; the best fit is \SI{2.1}{s}
later than the light-curve timing, but site and time are partly degenerate along the track (with the light-curve
timing the site moves by \SI{21}{km} and the rms rises from 11 to 18~arcmin), and we keep the light-curve timing.
Aperture photometry of the capsule in all three JAXA
videos ($r\le15$~px, G, linearised, background from an annulus at 25--40~px with the trail masked) shows that the head is
saturated in all frames of FireBall\_1 but one and in every frame of FireBall\_2, while the 45 unsaturated frames of
FireBall\_3 (of 1938) do not overlap the spectroscopic light curve enough for a scaling: scaled to $I_{400-680}$ at the faintest frames, the aperture flux falls short of $I$
near the peak (Figure~\ref{fig:train}b), so the brightness of the capsule in the videos is taken from the spectroscopic
light curve.

\appendix{Schematic temperature profile along the stagnation streamline}
\label{app:schematic}
The curves of Figure~\ref{fig:schematic}b are analytic shapes chosen to illustrate the structure of the shock layer;
they are not the result of a flow-field computation, and only the measured temperatures (the bands) are data.
The position is $x$, the distance from the wall in units of the shock standoff, with the shock at $x_s=0.97$ and
$s=x_s-x$ the distance behind the shock.
\begin{itemize}
\item Outside the boundary layer the translational temperature relaxes from a post-shock overshoot to a plateau,
and the vibrational/electronic temperature rises towards it:
\begin{equation*}
T_{\rm tr}=T_{\rm eq}+\Delta T\exp(-s/0.05),\qquad T_{\rm v}=T_{\rm eq}\left[1-\exp(-s/0.07)\right].
\end{equation*}
The plateau is set to $T_{\rm eq}=\SI{8800}{K}$, the upper
bound from the H$\beta$ Stark width and the Saha equation (Section~\ref{sec:lines}); it is not a computed
equilibrium temperature. The overshoot amplitude, $\Delta T=\SI{3.2e4}{K}$, is illustrative: the frozen post-shock
temperature from the Rankine--Hugoniot relations with $\gamma=1.4$, $T_2=2(\gamma-1)V^2/[(\gamma+1)^2R]$, is
$5.3$--$6.5\times10^4$~K at 10.4--\SI{11.6}{km\,s^{-1}}. The relaxation lengths (0.05 and 0.07 of the standoff)
are also illustrative.
\item In the boundary layer, $0\le x<\delta$ with $\delta=0.18$, both temperatures follow the cubic
(Pohlhausen-type) profile $T=T_w+(T_{\rm eq}-T_w)\left(\tfrac32\eta-\tfrac12\eta^3\right)$, $\eta=x/\delta$, from the
wall temperature $T_w=\SI{3000}{K}$, the colour temperature of the surface (Table~\ref{tab:temp}).
\item In the ablator ($x<0$) the temperature decreases as $T=T_w-\SI{2600}{K}\,[1-\exp(x/0.08)]$, a qualitative
thermal penetration.
\end{itemize}
The measured temperatures are drawn as bands over the regions to which they are assigned in the text: C$_2$ Swan
(7800--\SI{9200}{K}, 80--72~km) and CN (\SIrange{6500}{7500}{K}) over the boundary layer and the shock layer, the
Balmer excitation temperature (4300--\SI{6700}{K}) in and near the boundary layer, and the colour temperature of the
surface (2720--\SI{3080}{K}) at the wall. Their horizontal extent indicates only this assignment, since the
observations do not resolve the flow (Section~\ref{sec:limits}). The true standoff is about $0.07R_n$
(\SI{1.5}{cm}).

\appendix{Digest movie}
\label{app:movie}
A one-minute digest movie of the entry (hayabusa2\_digest.mp4; 3840$\times$2160 pixels, 30 frames per second,
62.6~s; English) is available in the Zenodo repository \citep{AbeZenodo}, together with an animation of the
emission features of Figure~\ref{fig:altspec} built up exposure by exposure with the light curve (h2\_spec\_anim\_en.mp4). After a title and the
annotated exposure of Figure~\ref{fig:frame}, it shows all 72 exposures with a capsule spectrum (79.7 to 49.2~km) at
half speed. Each frame gives the altitude, the time $t$, the apparent V magnitude and the colour temperature of the
exposure; the video strip of the first order and of the second order ($\times4$), made as in Figure~\ref{fig:app_strips};
the calibrated first-order spectrum at 400--\SI{670}{nm} with the grey-body fit; and the light curve, the
temperatures, the altitude and speed and the direction from Coober Pedy West, compared with a three-degree-of-freedom
trajectory model fitted to the observed directions and to the published pre-flight values \citep{Yamada2021}. This
model serves for illustration only; the altitudes and the analysis of this paper follow Section~\ref{sec:reduction}.
The movie ends with a summary of the results.

\bibliographystyle{sn-basic}
\bibliography{references}

\end{document}